\documentstyle[tighten,epsf,floats,aps,preprint]{revtex}
\preprint{KUNS 1281}
\newcommand{\argon}{{}^{40}\!{\rm Ar}}
\newcommand{\almi}{{}^{27}\!{\rm Al}}
\newcommand{\carbon}{{}^{12}{\rm C}}
\renewcommand{\vec}[1]{\bbox{#1}}
\newdimen\figsize \figsize=0.6\textwidth
\ifx\agt\undefined\def\agt{>\sim}\fi
\ifx\glt\undefined\fi
\begin{document}
\title{
Flow of Nucleons and Fragments in $\bbox{\argon+\almi}$ Collisions
Studied with AMD }
\author{
Akira Ono and Hisashi Horiuchi
}
\address{
Department of Physics, Kyoto University, Kyoto 606-01, Japan
}
\maketitle
\begin{abstract}
Collective transverse momentum flow of nucleons and fragments in
intermediate energy $\argon+\almi$ collisions is calculated with the
antisymmetrized molecular dynamics (AMD).  The observed flow and its
balance energy are reproduced very well by the calculation with the
Gogny force which corresponds to the soft EOS of the nuclear matter.
Especially the calculated absolute value of the fragment flow is
larger than that of the nucleon flow in the negative flow region,
which can be explained by the existence of two components of flow. In
addition to many similarities, difference in the deuteron flow is
found between $\carbon+\carbon$ and $\argon+\almi$ collisions, and its
origin is investigated by paying attention to the production mechanism
of light fragments. We also investigate the dependence of the flow of
nucleons and fragments on the stochastic collision cross section and
the effective interaction, and conclude that the stiff EOS without
momentum dependence of the mean field is not consistent to the
experimental data.
\end{abstract}

\narrowtext
\section{Introduction}
In intermediate and high energy heavy ion collisions, high density and
high temperature region is created, and the collective flow in heavy
ion reactions has been studied in order to extract the information on
the hot and dense nuclear matter.  Since the relevant energy scale
such as the compression energy per nucleon and the depth of the
nuclear mean field is several tens MeV, we can expect that
intermediate energy heavy ion collisions with the incident energy
region from several tens MeV/nucleon up to a few hundreds MeV/nucleon
are suitable for the above mentioned purpose.  At lower incident
energy in this energy region, the attractive (negative) flow pattern
is produced due to the attractive interaction between projectile and
target nuclei. And above a certain incident energy, called balance
energy, the flow changes into repulsive (positive) by the effects of
more two-nucleon collisions, higher compression and the momentum
dependence of the mean field.  This balance energy has been paid
attention to as a good indicator of the equation of state (EOS) of the
nuclear matter, but by the study with microscopic simulation
approaches such as Vlasov-Uehling-Uhlenbeck method, it has turned out
that extracting the EOS is not easy since the inclusive collective
flow reflects not only the EOS but also other factors such as the
cross section of the two-nucleon collision term which has theoretical
ambiguity in the nuclear medium.

Although the collective flow is usually considered as a one-body
observable, the most characteristic feature of intermediate energy
heavy ion reactions is the fragment formation.  Recently the
collective flow has become to be measured exclusively with the
identification of charges and/or mass numbers of fragments
\cite{PETER,PETERa,PETERb,WESTFALL,DOSS}.  The observed flow of
fragments has been found to be large compared to the flow of nucleons.
The reason is that most nucleons are emitted by the hard stochastic
collisions which erase largely the effect of the mean field in the
nucleon flow.  It suggests that the flow of composite fragments
carries precious information on the EOS (or the mean field).  In our
previous work on $\carbon+\carbon$ collisions, we found that the flow
in the incident energy region of negative flow consists of two
components at the end of the dynamical stage of the reaction, i.e.,
the flow of nucleons emitted by stochastic collisions and the flow of
excited fragments or the nuclear matter which is largely affected by
the mean field.  Thus we can expect that the systematic study of the
fragment flow together with the nucleon flow may give us important
information on the EOS.

For the theoretical analysis of the flow of fragments, the model
should be able to describe the dynamical fragment formation, and we
use the antisymmetrized molecular dynamics with stochastic collisions
(AMD) \cite{ONOa,ONOb,ONOc,ONOd,FELD,HORI,HORIa}.  In AMD, the system
is described with a Slater determinant of Gaussian wave packets and
the time development of the centers of the wave packets is determined
by the time dependent variational principle and the stochastic
collision process. We have demonstrated that the AMD can describe some
quantum mechanical features such as the shell effect in the dynamical
production of fragments in the intermediate energy heavy ion
reactions. Furthermore, since we use the antisymmetrized wave function
in AMD, the ground states of initial nuclei are the most precisely
described among many simulation methods for heavy ion reactions and
have no ambiguity in the choice of them because the ground states are
the states which give the minimum value of the Hamiltonian with an
effective interaction. By using a finite range effective interaction
such as the Gogny force, the momentum dependence of the mean field is
automatically taken into account.

In this paper, we calculate $\argon+\almi$ collisions with AMD in the
incident energy region 25 MeV $\le E/A\le$ 135 MeV and analyze the
collective transverse momentum flow of nucleons and fragments as well
as other features of their momentum distribution. The purpose of this
paper is two fold. One is to get understanding on the production
mechanism of the flow of nucleons and fragments.  It is very important
to understand the reaction mechanism before getting anything from
heavy ion reactions. We will really see that the production mechanism
of light fragments such as deuterons and $\alpha$ particles is closely
related to their flow by comparing the present results with our
previous results for $\carbon+\carbon$ collisions \cite{ONOd}. The
second purpose is to investigate the dependence of the flow on the
two-nucleon collision cross section and on the effective interaction,
based on which we discuss the determination of the EOS of the nuclear
matter. We make calculations with Gogny force which gives the soft EOS
of the nuclear matter and momentum dependent mean field and with a
kind of Skyrme force (called SKG2 in this paper) which gives stiff EOS
and mean field with little momentum dependence. It will be concluded
that the Gogny force reproduce the observed data very well while the
SKG2 force does not, in spite of some ambiguity of the stochastic
collision cross section.

This paper is organized as follows. After explaining the outline of
the framework briefly in Sec.\ II, some calculated quantities
such as mass distribution and momentum distribution of produced
fragments are presented in Sec.\ III in order to show the
overall features of the reaction investigated in this paper. In Sec.\
IV, the calculated results of the flow of nucleons and 
fragments with the Gogny force are discussed, and in Sec.\ V
the flow of light fragments such as deuterons is investigated by
paying attention to their production mechanism and comparing it with
the case of $\carbon+\carbon$ collisions. The dependence of the
results on the stochastic collision cross section and on the adopted
effective interaction is discussed in Sec.\ VI in order to
determine the EOS. Section VII is devoted to the summary.

\section{Outline of the Method}

Since the framework of the antisymmetrized version of molecular
dynamics (AMD) was described in detail in Refs.\
\CITE{ONOb,ONOc,ONOd}, here is shown only the outline of our
framework.

In AMD, the wave function of $A$-nucleon system is described by a
Slater determinant $|\Phi(Z)\rangle$,
\begin{equation}
  |\Phi(Z)\rangle=
   {1\over\sqrt{A!}}\det\Bigl[\varphi_i(j)\Bigr],\quad
   \varphi_i=\phi_{{\vec Z}_i}\chi_{\alpha_i},
\end{equation}
where $\alpha_i$ represents the spin-isospin label of $i$th single
particle state, $\alpha_i={\rm p}\uparrow$, ${\rm p}\downarrow$, ${\rm
n}\uparrow$, or ${\rm n}\downarrow$, and $\chi$ is the spin-isospin
wave function. $\phi_{{\vec Z}_i}$ is the spatial wave function of
$i$th single particle state which is a Gaussian wave packet
\begin{equation}
  \langle{\vec r}\,|\phi_{{\vec Z}_i}\rangle=
  \Bigl({2\nu\over\pi}\Bigr)^{3/4}
  \exp\biggl[-\nu\Bigl({\vec r}-{{\vec Z}_{i}\over\sqrt\nu}\Bigr)^2
                      +{\textstyle{1\over2}}{{\vec Z}}_{i}^2\biggr],
\end{equation}
where the width parameter $\nu$ is treated as time-independent in our
model.  We took $\nu=0.16$ ${\rm fm}^{-2}$ in the calculation
presented in this paper.

The time developments of the centers of Gaussian wave packets,
$Z=\{{\vec Z}_i\;(i=1,2,\ldots,A)\}$, are determined by two processes.
One is the time development determined by the time-dependent
variational principle
\begin{equation}
  \delta\int_{t_1}^{t_2}dt\,
  { \langle\Phi(Z)|(i\hbar{d\over dt}-H)|\Phi(Z)\rangle
   \over\langle\Phi(Z)|\Phi(Z)\rangle}=0,
\end{equation}
which leads to the equation of motion for $Z$.

The second process which determines the time development of the system
is the stochastic collision process due to the residual interaction.
We incorporate this process in the similar way to the quantum
molecular dynamics (QMD) by introducing the physical coordinates
$\{{\vec W}_i\}$ \cite{ONOa,ONOb} as
\begin{equation}
  {\vec W}_i=\sum_{j=1}^A \Bigl(\sqrt Q\Bigr)_{ij}{\vec Z}_j,
\quad
  Q_{ij}=
\ifpreprintsty
         {\partial\over\partial({\vec Z}_i^*\cdot{\vec Z}_j)}
         \log \langle\Phi(Z)|\Phi(Z)\rangle.
\else
         {\partial\log \langle\Phi(Z)|\Phi(Z)\rangle
         \over\partial({\vec Z}_i^*\cdot{\vec Z}_j)}.
\fi
\label{eq:physcoord}
\end{equation}
We use the same energy and density dependent two-nucleon collision
cross section as in our previous work \cite{ONOd} which is based on
the free cross section of $pp$ and $pn$ scattering and has some
reduction in the nuclear medium. The detail is explained in the
Appendix B of Ref.\ \CITE{ONOd}. Nucleon-alpha collisions are switched
off in the calculation of $\argon+\almi$ collisions in this paper.

The simulations of AMD are truncated at a finite time $t=t_{\rm sw}$
($t_{\rm sw}=150$ fm/$c$ in most calculations presented in this paper
and $t_{\rm sw}=225$ fm/$c$ when incident energy is lower than 30
MeV/nucleon).  The dynamical stage of the reaction has finished by
this time and some excited fragments have been formed which will emit
lighter particles with a long time scale.  Such statistical decays of
the equilibrated fragments are calculated with a code of Ref.\
\CITE{MARUb} which is similar to the code of P\"uhlhofer
\cite{PUHLHOFER}.

One of the most important inputs in the study of the collective flow
is the choice of the effective interaction. The required property of
the effective interaction for the study of flow is that it should
reproduce the saturation of the nuclear matter and the bulk properties
of nuclei in wide mass number region within the framework of AMD. In
this paper we execute calculations with two effective interactions
which give different stiffness of the nuclear matter, aiming to
determine the EOS by the comparison to the data of flow. The first one
is the Gogny force \cite{GOGNY} which is composed of the finite range
two-body force and the density dependent zero range repulsive force. 
This force gives a soft EOS of the nuclear matter with the
incompressibility $K=228$ MeV, and also it gives momentum dependent
mean field which reproduce the observed energy dependence of the
nucleon optical potential depth up to the incident energy 200
MeV/nucleon. The second effective interaction used here is a
Skyrme-type interaction. Just for the convenience of the numerical
calculation we use a modified version of the effective interaction
used in Ref.\ \CITE{OHNISHI-WADA}, which has the form
\ifpreprintsty 
\begin{eqnarray}
v({\vec r}_i,{\vec r}_j)
 &=&v_0((1-m)-mP_\sigma P_\tau)
    \exp\bigl[-({\vec r}_i-{\vec r}_j)^2/\mu^2\bigr]
\nonumber\\
 &+&{t_\rho\over6}
    (W_\rho+B_\rho P_\sigma-H_\rho P_\tau-M_\rho P_\sigma P_\tau)
               \rho({\vec r}_i)\delta({\vec r}_i-{\vec r}_j),
\label{eq:skg2force}
\end{eqnarray}
\else 
\begin{eqnarray}
v({\vec r}_i,{\vec r}_j)
 &=&v_0((1-m)-mP_\sigma P_\tau)
    \exp\bigl[-({\vec r}_i-{\vec r}_j)^2/\mu^2\bigr]
\nonumber\\
 &+&{t_\rho\over6}
    (W_\rho+B_\rho P_\sigma-H_\rho P_\tau-M_\rho P_\sigma P_\tau)
\nonumber\\
 &&\qquad\times\rho({\vec r}_i)\delta({\vec r}_i-{\vec r}_j),
\label{eq:skg2force}
\end{eqnarray}
\fi 
where $P_\sigma$ and $P_\tau$ are the spin and isospin exchange
operators respectively.  We call this interaction SKG2 force and the
parameters are listed in Table \ref{table:skg2force}. This force gives
similar property of Skyrme VII force; namely stiff EOS of nuclear
matter with the incompressibility $K=373$ MeV and no momentum
dependence of the mean field. We have modified the density dependent
part from Ref.\ \CITE{OHNISHI-WADA} so that the reasonable symmetry
energy is obtained. Without this modification, we found that many
neutron-rich fragments are produced unphysically in heavy ion
collisions.

\begin{table}
\caption{\label{table:skg2force}
Parameters of SKG2 force in Eq.\ (\protect\ref{eq:skg2force}).}
\begin{tabular}{cccccccc}
$v_0$ [MeV] & $m$ & $\mu$ [fm] & $t_\rho$ [MeV fm$^6$]
& $W_\rho$ & $B_\rho$ & $H_\rho$ & $M_\rho$ \\
\hline
$-624.46$   & 0.2 & 0.68     & 17269.8    & 1.0 & 0.2 &$-0.8$ & 0.0 \\
\end{tabular}
\end{table}

Since in AMD the center-of-mass motion of the fragment is described by
a Gaussian wave packet, we have to subtract from the AMD Hamiltonian
$\langle\Phi(Z)|H|\Phi(Z)\rangle/\langle\Phi(Z)|\Phi(Z)\rangle$ the
sum of the spurious zero-point energies of the fragments whose number
changes with time \cite{ONOb}.  The prescription to deal with this
problem is given in the appendix C of Ref.\ \CITE{ONOd}, and the
adopted parameters which are used with Gogny force and SKG2 force are
listed in Table \ref{table:subzero}.

\begin{table}
\caption{\label{table:subzero}
Parameters concerned with the subtraction of spurious zero-point
oscillations of fragments.  Parameters which are used with the Gogny
force and SKG2 force in this paper are shown. See Appendix C of Ref.\
\protect\CITE{ONOd} for detail.
}
\begin{tabular}{ccccccccccc}
Force & $\xi$ & $a$ & $\hat\xi$ & $\hat a$ & $\bar\xi$ & $\bar a$
& $g_0$ & $\sigma$ & $M$ & $T_0$ [MeV] \\
\tableline
Gogny  &2.0&0.6&2.0&0.2&1.0&0.5&1.0&2.0&12.0&9.2\\
SKG2   &2.0&0.6&2.0&0.2&---&---&---&---&---&9.0\\
\end{tabular}
\end{table}

In Table \ref{table:groundstates}, we show the the ground states of
$\argon$ and $\almi$ nuclei which are obtained by the frictional
cooling method in AMD and used as the initial states of AMD
simulations. Both the Gogny force and the SKG2 force give similar
ground states with appropriate binding energies. The ground state of
$\argon$ is almost spherical but the obtained $\almi$ ground state is
prolately deformed with $\beta\sim0.4$. Here the deformation
parameters $\beta$ and $\gamma$ are defined by
\begin{equation}
t_i=\sqrt{5\over4\pi}\,\beta
    \cos\Bigl(\gamma-{2\pi\over3}i\Bigr)
\qquad\hbox{for $i=1,2,3$},
\label{eq:defbeta}
\end{equation}
where $e^{2t_1}:e^{2t_2}:e^{2t_3}$ is the ratio of eigenvalues of the
inertia tensor $\langle r_i r_j\rangle$ with $t_1+t_2+t_3=0$ and
$t_1\le t_2\le t_3$.

\begin{table}
\caption{\label{table:groundstates}
Properties of ground states which are obtained by the frictional
cooling method and used as the initial states of AMD simulations.
There are shown binding energy, root mean square radius, quadrapole
deformation parameters $\beta$ and $\gamma$. }
\begin{tabular}{cccccc}
Nucleus & Force &
 B.E. [MeV] & $\sqrt{\langle r^2\rangle}$ [fm] & $\beta$ & $\gamma$ \\
\hline
$\argon$  & Gogny & 335        &  3.35     & 0.16 & $<10^\circ$\\
          & SKG2  & 340        &  3.36     & 0.09 & $35^\circ$\\
          & exp.  & 344        &           &      &    \\
\hline
$\almi$   & Gogny & 220        &  3.17     & 0.40 & $<2^\circ$\\
          & SKG2  & 225        &  3.19     & 0.42 & $<2^\circ$\\
          & exp.  & 225        &           &      &    \\
\end{tabular}
\end{table}

\section{Overall Features of the Reactions}

We have calculated the $\argon+\almi$ collisions with the incident
energy 25 MeV $\le E/A \le$ 135 MeV. In order to see the overall
features of the reaction in this energy region, we will show some
calculated quantities with Gogny force in this section.

\begin{figure}
\ifx\epsfbox\undefined\else
\centerline{\epsfxsize=\figsize\epsfbox{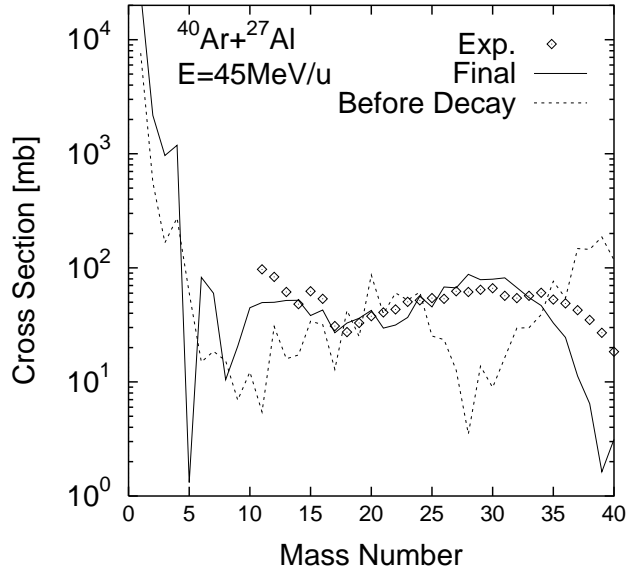}}
\fi
\caption{\label{figure:ArAl45-msdst}
Mass distribution of fragments in $\argon+\almi$ reaction at 45
MeV/nucleon.  Calculated result at the end of the AMD simulation
before the statistical decay is shown by dashed line as well as the
final calculated result by solid line. Gogny force was used as the
effective interaction. Fragments which are emitted with the angle
$\theta > 5^\circ$ and the energy $E/A>3$ MeV are taken into
calculation. Diamonds are the experimental data by Dayras et al.
\protect\cite{DAYRAS} for $\argon+\almi$ at 44 MeV/nucleon.  }
\end{figure}

Figure \ref{figure:ArAl45-msdst} shows the calculated mass
distribution of produced fragments at the incident energy 45
MeV/nucleon compared with the experimental data of $\argon+\almi$
collisions at 44 MeV/nucleon by Dayras et al. \cite{DAYRAS}. Dashed
line shows the mass distribution at $t=t_{\rm sw}$, namely the result
of AMD simulation before the statistical cascade decay, and the solid
line shows the final result after the calculation of the statistical
cascade decay. In the calculation, we have included only the fragments
which are emitted with angles $\theta>5^\circ$ in the laboratory
frame, and the calculated result depends on this threshold angle in
the mass number region $A\agt25$. It can be seen that the observed
mass distribution is reproduced as the results of the statistical
decay of projectile-like and target-like fragments.  It should be
noted that the shell effect in the dynamical stage of the reaction has
appeared in the large yield of $A=4$ fragment before the statistical
decay but the statistical accuracy is not sufficient in heavier mass
number region to discuss the shell effect. The present result of AMD
is quite similar to the result of QMD in Ref.\ \CITE{MARUb} in the
mass number region $A>5$.

\begin{figure}
\widetext
\ifx\epsfbox\undefined\else
\epsfxsize=0.5\textwidth
\centerline{
\begin{minipage}{0.5\textwidth}\epsfbox{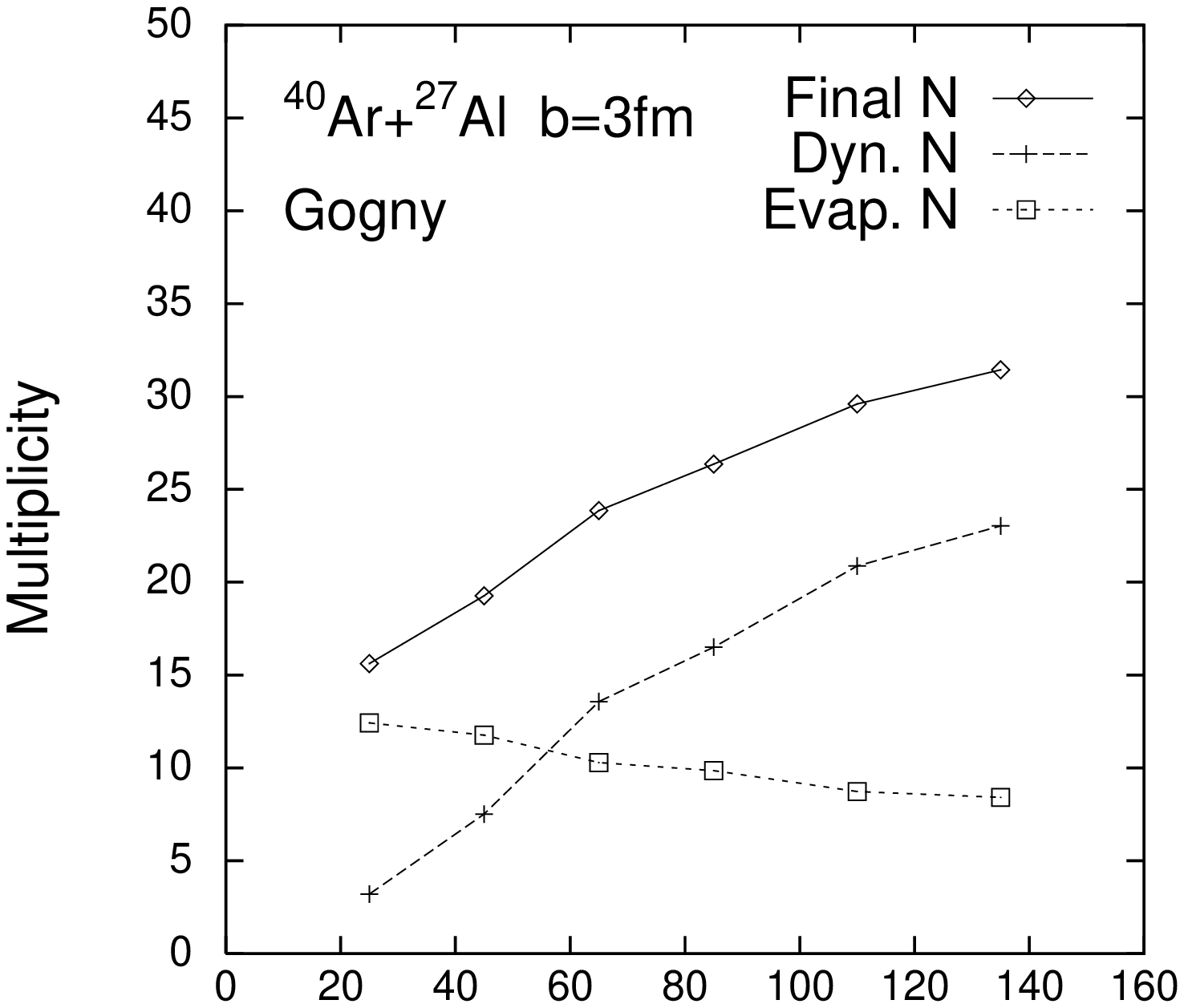}\end{minipage}
\hspace{-0.14\textwidth}
\begin{minipage}{0.5\textwidth}\epsfbox{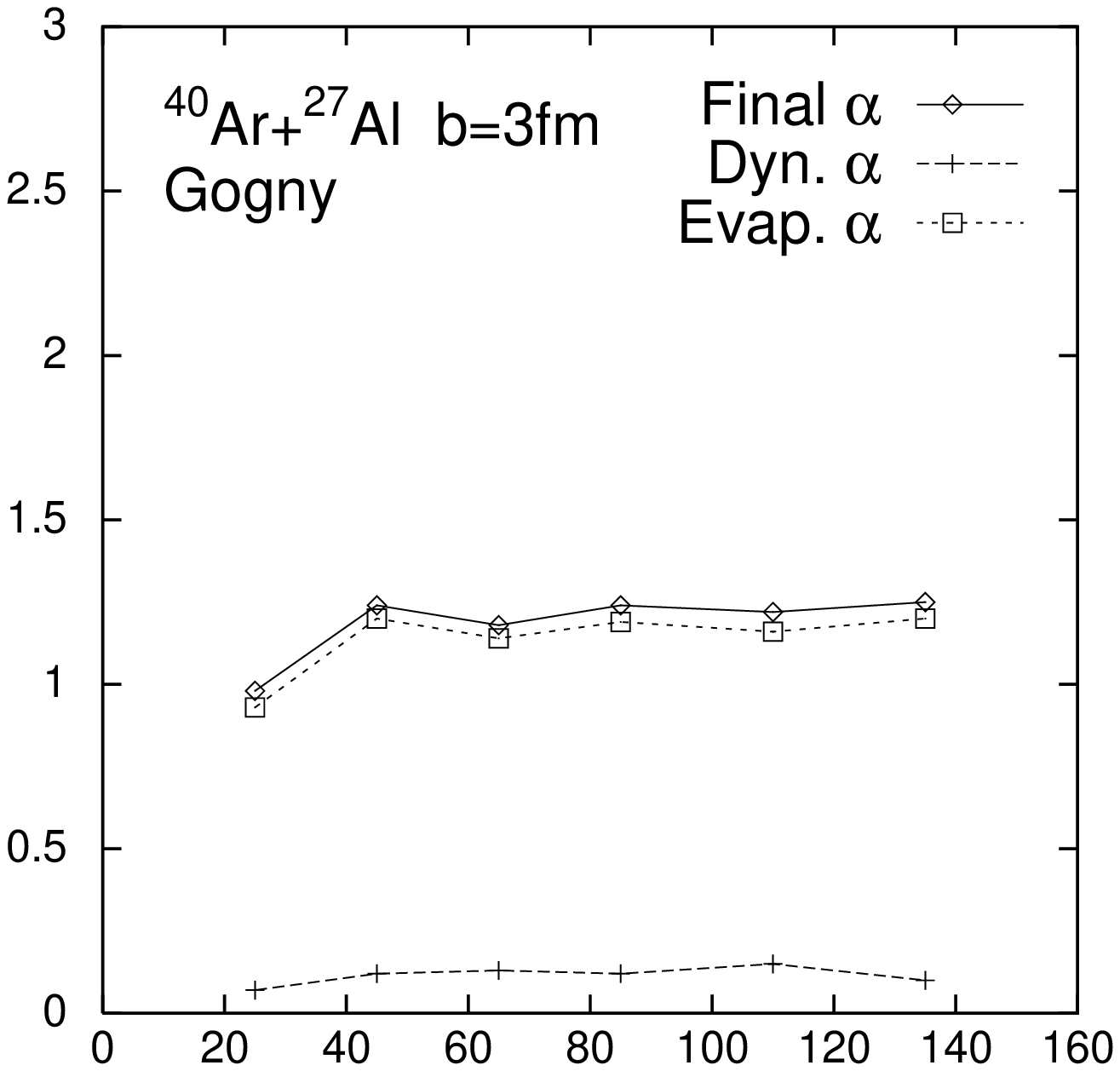}\end{minipage}
}
\vspace{-0.08\textwidth}
\centerline{
\begin{minipage}{0.5\textwidth}\epsfbox{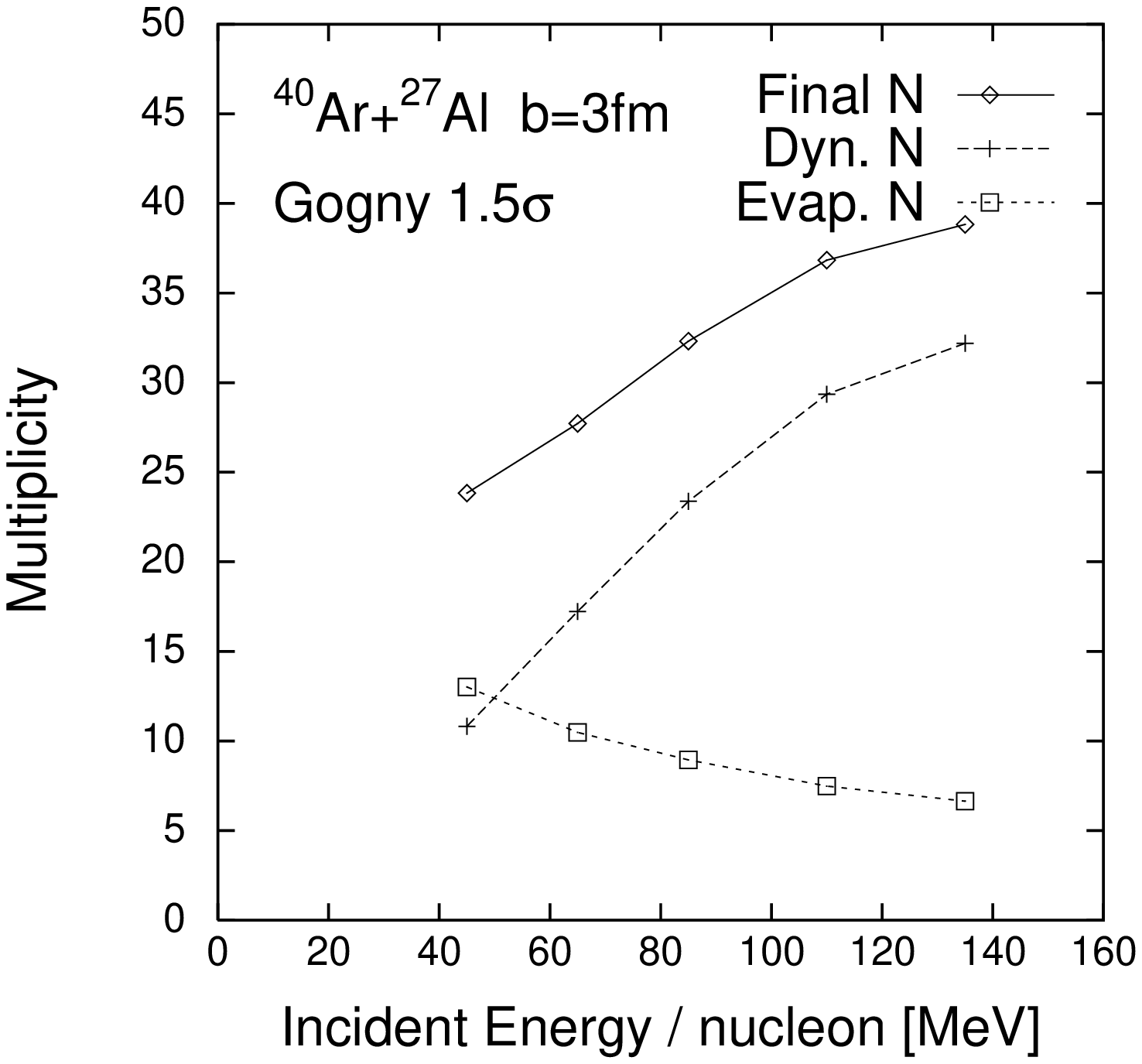}\end{minipage}
\hspace{-0.14\textwidth}
\begin{minipage}{0.5\textwidth}\epsfbox{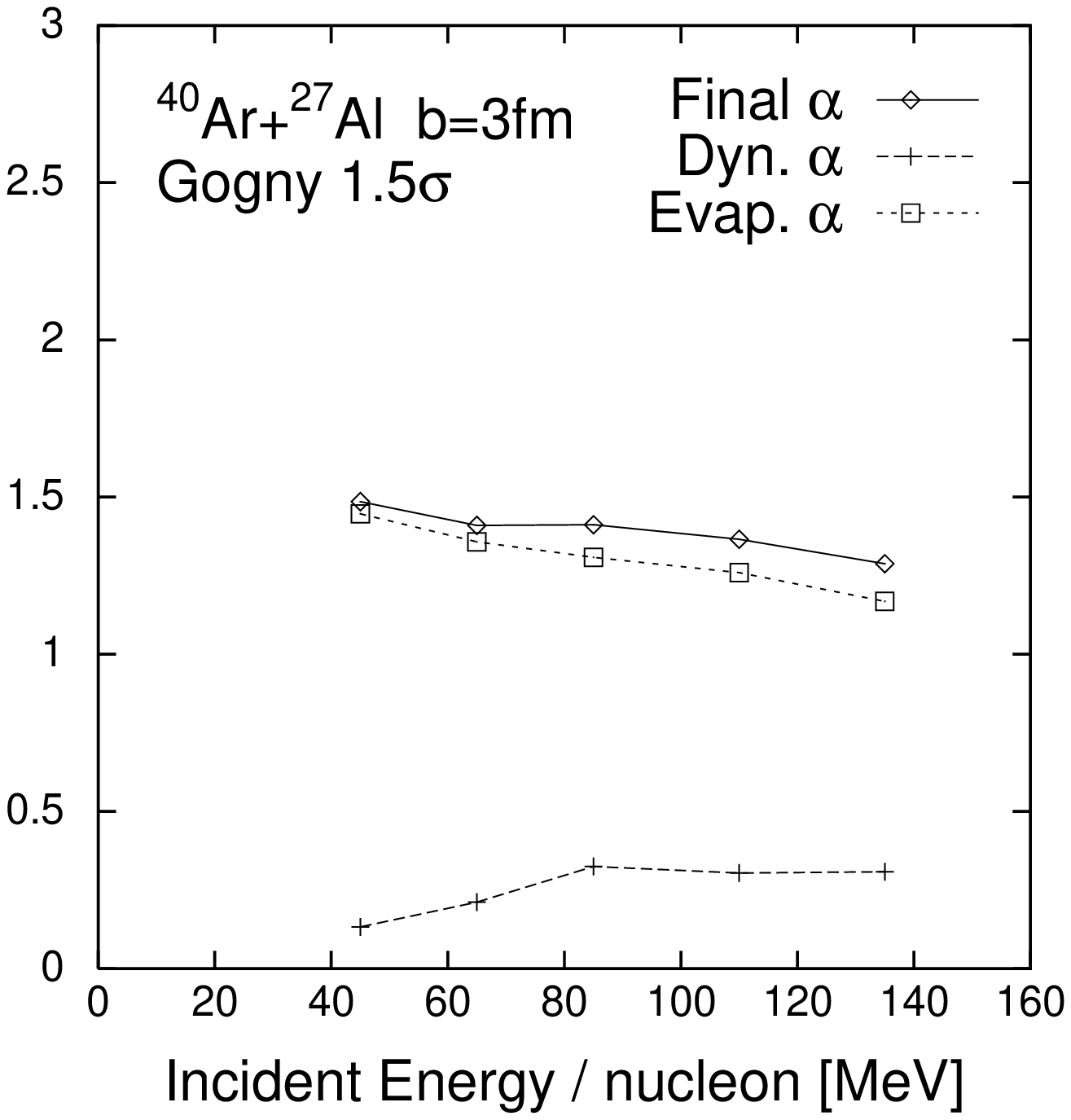}\end{minipage}
}
\fi
\caption{\label{figure:mlteng}
Calculated multiplicities of nucleons (left) and $\alpha$ particles
(right) in $\argon+\almi$ collisions with $b=3$ fm as functions of the
incident energy. The multiplicities of dynamically produced particles
(pluses) which existed at $t=t_{\rm sw}$ and the multiplicities of
particles produced in the statistical decay process (squares) are
shown as well as the final result (diamonds). Lower two figures are
the results of the calculation with the stochastic two-nucleon
collision cross section $1.5\sigma$.}
\narrowtext
\end{figure}

In Fig.\ \ref{figure:mlteng}, the multiplicities of nucleons and
$\alpha$ particles are shown as functions of the incident energy.  We
fix the impact parameter to 3 fm here and in the following.  In
addition to the final multiplicities, there are shown the
multiplicities of dynamically produced particles which existed at
$t=t_{\rm sw}$ and the multiplicities of evaporated particles in the
statistical decay process.  One should keep in mind that the dynamical
stage of the reaction has already finished at $t=t_{\rm sw}$, and
therefore the `dynamically produced' particles defined here include
some particles {\it evaporated} from excited fragments which have
reached the equilibration before $t=t_{\rm sw}$. However the main part
of the `dynamically produced' particles have been produced before the
equilibration. In the calculation of lower two figures, we have
increased the stochastic two-nucleon collision cross section by a
factor 1.5 (denoted by $1.5\sigma$ for simplicity) compared to the
calculation of upper two figures where we use the standard cross
section ($1.0\sigma$) parameterized in the Appendix B of Ref.\
\CITE{ONOd}.  As can be expected, the nucleon multiplicity, especially
the multiplicity of dynamically emitted nucleons, increases as the
incident energy.  With $1.0\sigma$, about a quarter of total nucleons
are emitted as single nucleons at 25 MeV/nucleon, while this ratio is
about a half at 135 MeV/nucleon. On the other hand, the multiplicity
of $\alpha$ particles is almost independent of the incident energy in
the energy region investigated here. The features of nucleon
multiplicity are similar to what we observed in the calculation
$\carbon+\carbon$ in Ref.\ \CITE{ONOd}.  On the contrary, the
multiplicity of dynamically produced $\alpha$ particle is much smaller
than in $\carbon+\carbon$ collisions with $b=2$ fm where it was about
one dynamical $\alpha$ particle per event. This suggests that light
nuclei break more easily into small fragments than heavy nuclei. As
the two-nucleon collision cross section is increased from $1.0\sigma$
to $1.5\sigma$, the dynamical $\alpha$ multiplicity increases as well
as the dynamical nucleon multiplicity.

\begin{figure}
\widetext
\ifx\epsfbox\undefined\else
\epsfxsize=0.5\textwidth
\centerline{
\begin{minipage}{0.5\textwidth}\epsfbox{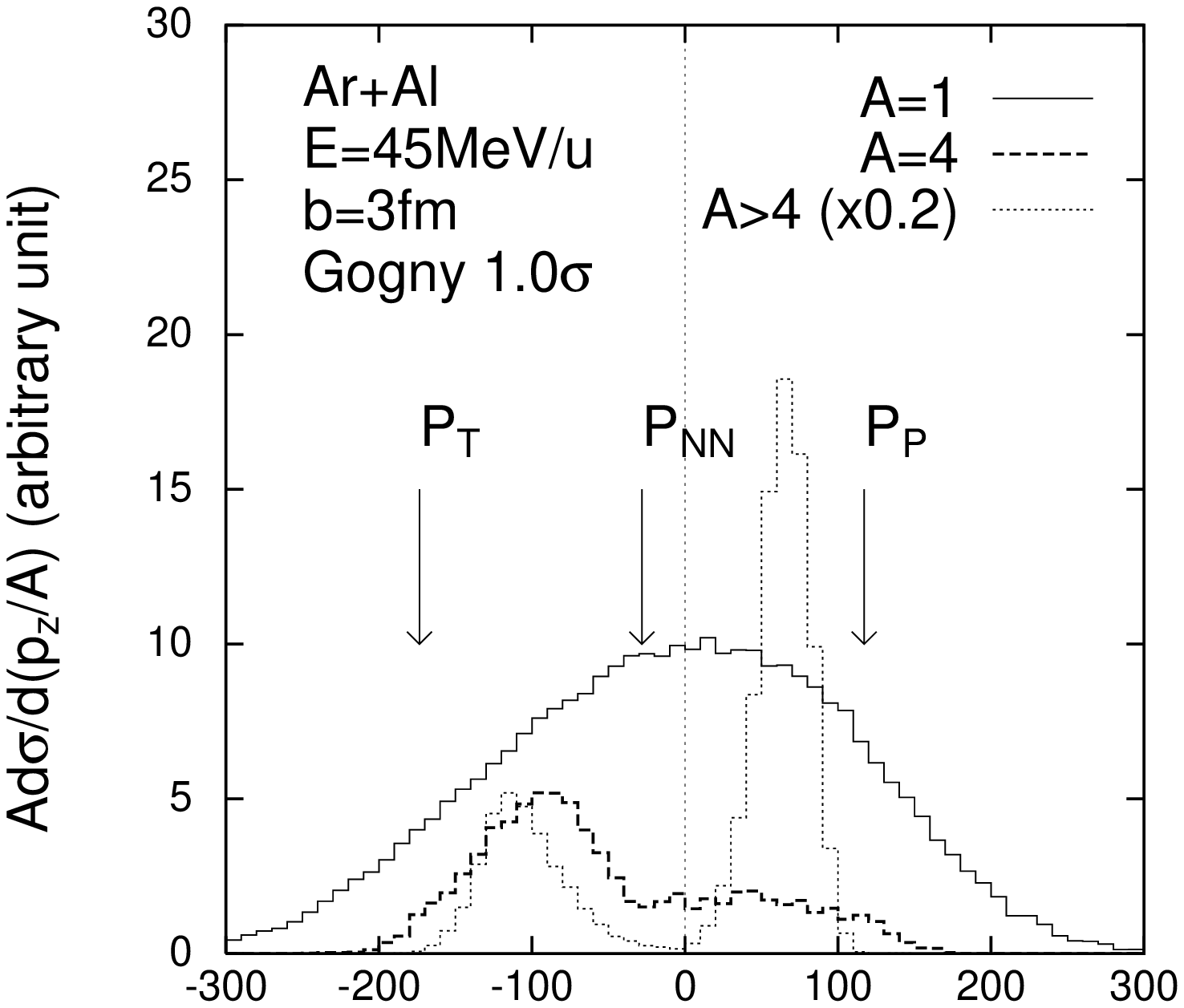}\end{minipage}
\hspace{-0.14\textwidth}
\begin{minipage}{0.5\textwidth}\epsfbox{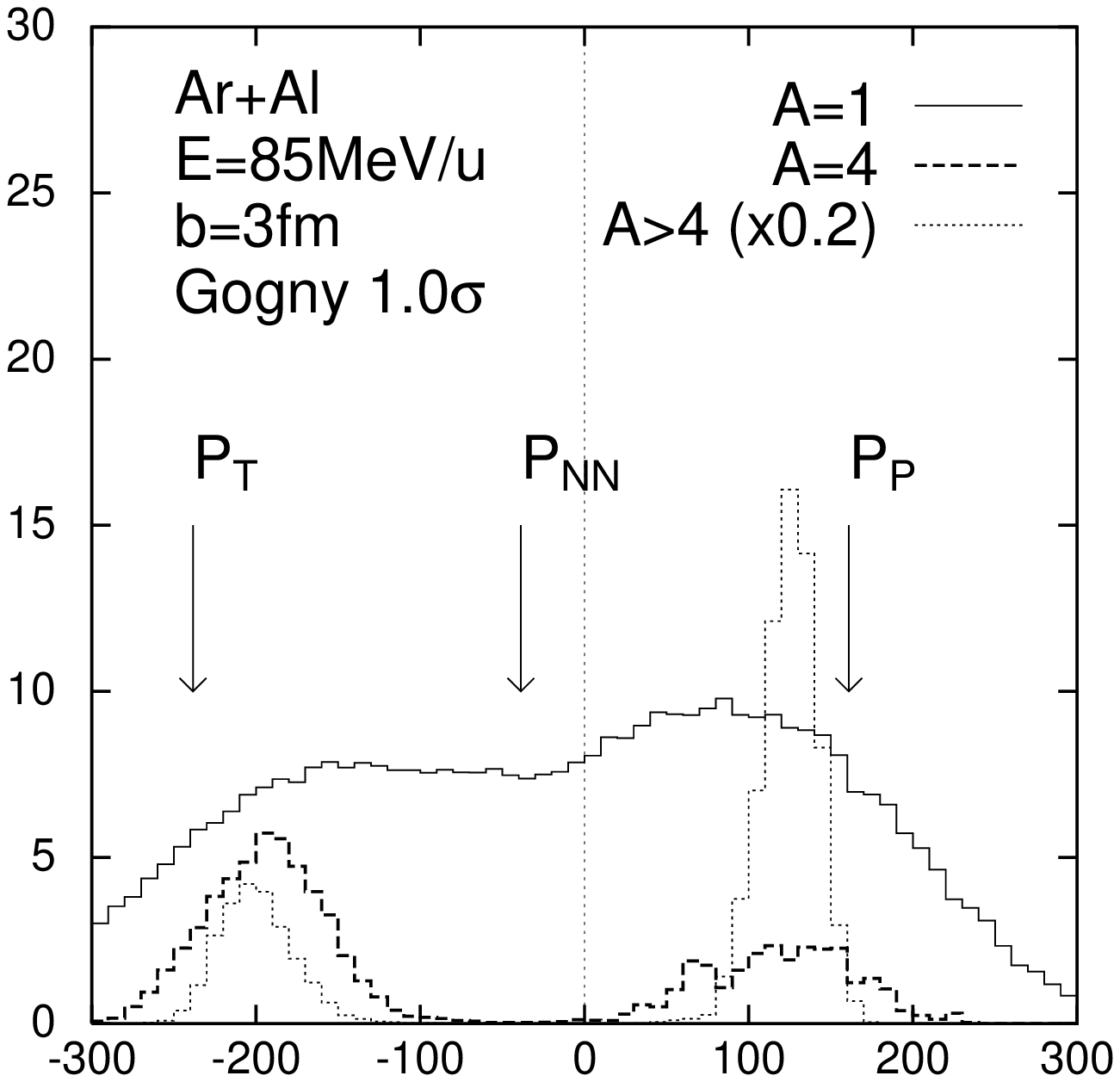}\end{minipage}
}
\vspace{-0.08\textwidth}
\centerline{
\begin{minipage}{0.5\textwidth}\epsfbox{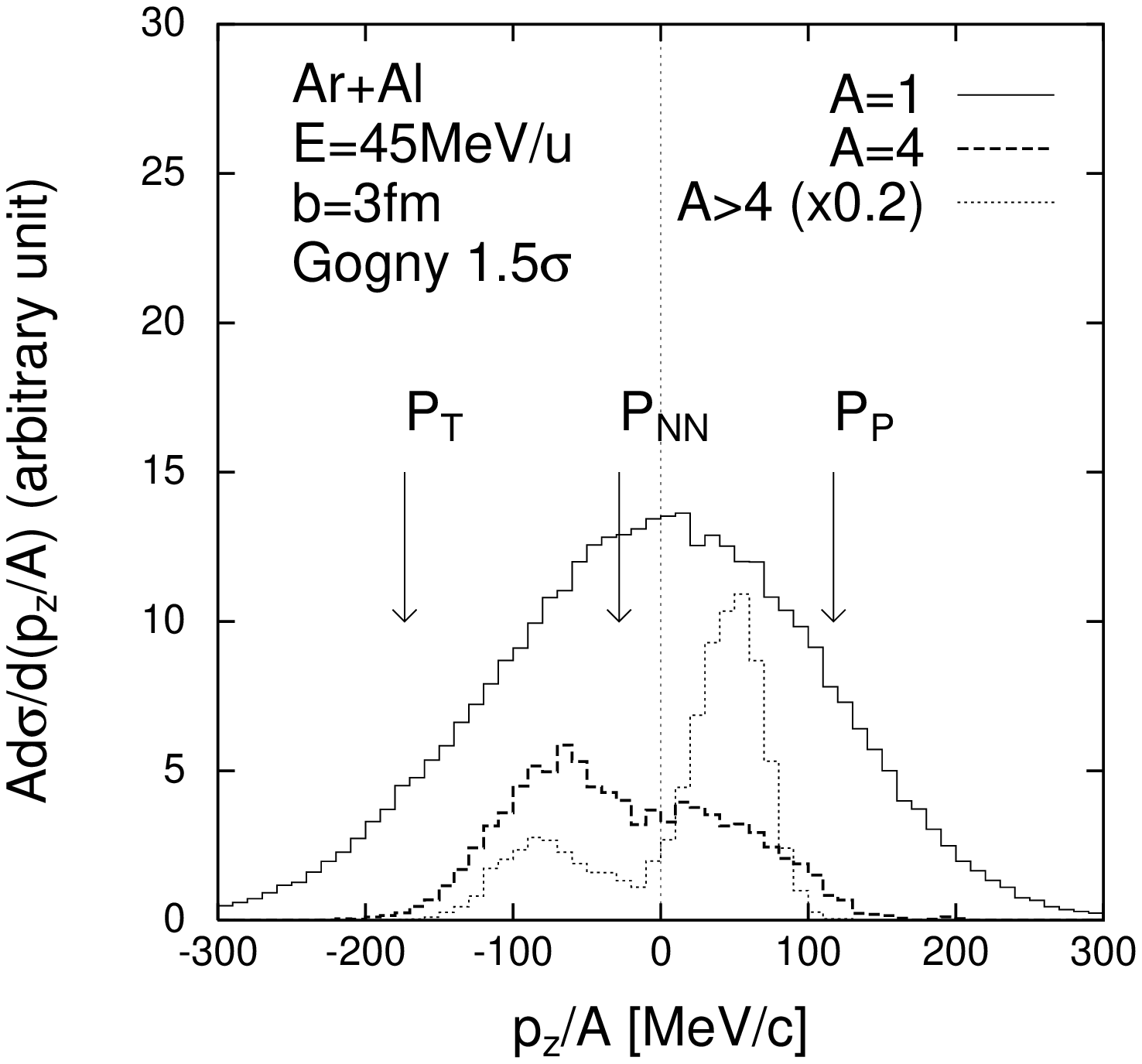}\end{minipage}
\hspace{-0.14\textwidth}
\begin{minipage}{0.5\textwidth}\epsfbox{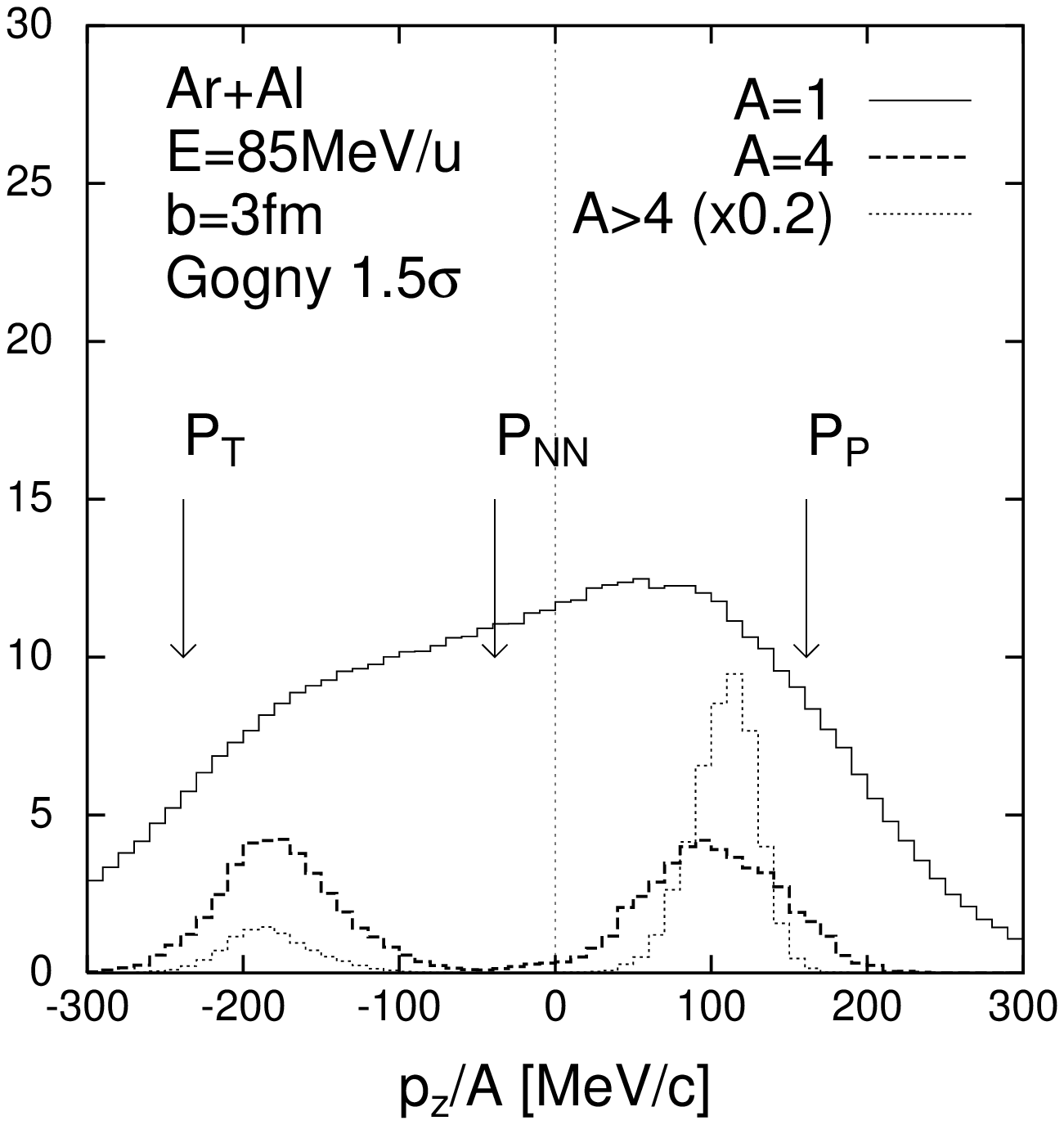}\end{minipage}
}
\fi
\caption{\label{figure:pzdst}
Calculated parallel momentum distribution of nucleons and fragments in
$\argon+\almi$ collisions with the incident energy 45 MeV/nucleon
(left) and 85 MeV/nucleon (right). The impact parameter is 3 fm. Gogny
force was used and the stochastic two-nucleon collision cross section
is $1.0\sigma$ (upper) and $1.5\sigma$ (lower). Vertical scale in
arbitrary unit is proportional to the number of nucleons contained in
the fragments. $p_z$ is the momentum component along the beam axis in
the center of mass frame. $P_P$, $P_T$ and $P_{NN}$ indicated by
arrows correspond respectively to the projectile velocity, target
velocity and the center-of-mass velocity of a projectile nucleon and a
target nucleon.}
\narrowtext
\end{figure}

In Fig.\ \ref{figure:pzdst}, the parallel momentum spectra of nucleons
and fragments in the collisions with the impact parameter 3 fm are
shown for two incident energies 45 MeV/nucleon and 85 MeV/nucleon. The
results with $1.0\sigma$ and $1.5\sigma$ are shown in upper part and
in lower part, respectively. Since the center-of-mass motions of
fragments are described with wave packets of Gaussian form with the
standard deviation $\hbar\sqrt{\nu}/\sqrt{A_F}$ in their momenta per
nucleon, we have assumed that the fragments with mass number $A_F$
produced by AMD simulations before the statistical decay have some
width in the momentum distribution. Taking account of the fact that
part of the width comes from the unphysical width of the initial
momenta of the projectile and the target especially for heavy
fragments \cite{ONOd}, we have attributed the width
$\hbar\sqrt{\nu}/A_F$ to their momenta per nucleon. It is clearly seen
that there are projectile-like and target-like components in the
momentum distribution of heavy fragments, while there is a large
component in the nucleon spectrum centered around the center-of-mass
velocity of two nucleons in projectile and target ($P_{NN}$ in
figures) or the total center-of-mass velocity. As for $\alpha$
particles, there is some yield around the center-of-mass velocity at
45 MeV/nucleon but projectile-like and target-like components clearly
separate from each other at 85 MeV/nucleon without any component left
around the center-of-mass velocity.  More $\alpha$ particles and less
heavier fragments are produced in target-like momentum region than in
projectile-like momentum region, which means the target nucleus
$\almi$ breaks up into smaller pieces compared to the projectile
nucleus $\argon$. The effect of the change of the stochastic collision
cross section can be seen in the these spectra. In addition to the
yield of nucleons and fragments, the peak position of projectile-like
and target-like components are sensitive to the cross section. The
peak position of the projectile-like components calculated with the
stochastic collision cross section around the investigated value
$1.0\sigma$ or $1.5\sigma$ seems consistent to the experimental data
\cite{PETERa}, but the detailed comparison is difficult here due to
the treatment of the impact parameter determination.

\section{Calculated Results of Flow}

The collective transverse momentum flow is an observable quantity
which reflects the interaction during the heavy ion collisions more
sensitively than the quantity discussed in the previous section. In
this paper, the flow is defined by
\begin{equation}
\langle wP_x/A\rangle
= {\sum_k A_k \mathop{\rm sign}(P_{kz})P_{kx}/A_k \over \sum_k A_k},
\label{eq:flow2def}
\end{equation}
where $k$ is the index of the produced fragments in all events and
$A_k$ are their mass numbers. $P_{kz}$ and $P_{kx}$ are the components
of the fragment momenta in the center-of-mass system along the beam
direction and the transverse direction in the reaction plane
respectively.  The direction of $x$ axis is taken so that the positive
value of flow means repulsive flow. When summation is taken over all
nucleons and fragments in Eq.\ (\ref{eq:flow2def}), it means the
inclusive flow. In the following we limit the summation to the
fragments with a specific mass number and discuss the exclusive flow.

It is also possible to use the more commonly adopted definition of
flow as the slope parameter of $\langle P_x/A\rangle$-$V_z$ curve
\cite{PETER,PETERb}
\begin{equation}
{V_P-V_T\over 2}
\left.{d\langle P_x/A\rangle \over dV_z}
\right|_{V_z=(V_P+V_T)/2},
\label{eq:flowdef}
\end{equation}
where $V_P$ and $V_T$ are the projectile and target velocities
respectively. However it is more convenient in our numerical
calculation to adopt the definition Eq.\ (\ref{eq:flow2def}) because
the statistical error is smaller than when Eq.\ (\ref{eq:flowdef}) is
adopted.  Furthermore the flow $\langle wP_x/A\rangle$ in Eq.\
(\ref{eq:flow2def}) is almost free from the ambiguity in the momentum
width of the produced fragments in AMD simulations. We have checked in
our previous work \cite{ONOd} that both definitions of the flow give
similar results except for the difference in the absolute value.

\begin{figure}
\ifx\epsfbox\undefined\else
\epsfxsize=\figsize
\centerline{\epsfbox{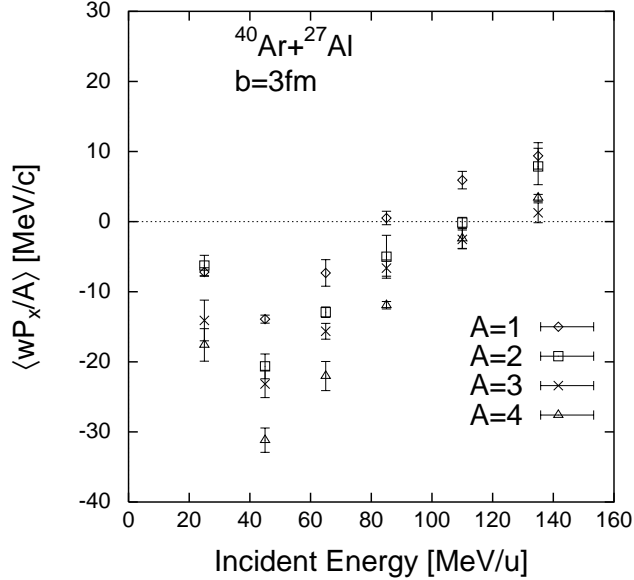}}
\fi
\caption{\label{figure:flow2}
Calculated results of the flow of nucleons and fragments with mass
number $A=1,2,3,4$ in the $\argon+\almi$ collisions at various
incident energies. The impact parameter is fixed to be 3 fm. Error
bars show estimated statistical error of the calculation.  }
\end{figure}

\begin{figure}
\ifx\epsfbox\undefined\else
\epsfxsize=\figsize
\centerline{\epsfbox{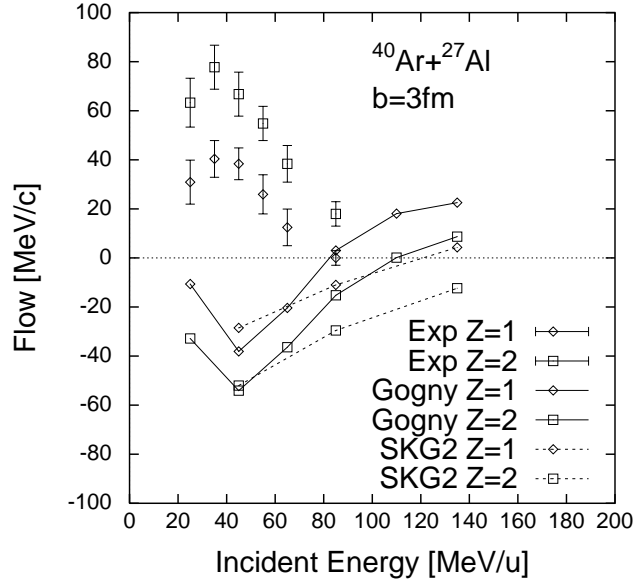}}
\fi
\caption{\label{figure:flowcomp}
Comparison of the calculated results with the experimental data
\protect\cite{PETERb}. Calculated flow in this figure is defined
as the slope parameter of $\langle P_x/A\rangle$-$V_z$ curve in the
similar way to Refs.\ \protect\CITE{PETER,PETERb}.  Shown experimental
data is the data in Fig.\ 3 of Ref.\ \protect\CITE{PETERb} multiplied
by a factor introduced in Ref.\ \protect\CITE{PETER} for the
correction of the reaction plane determination. The experimental data
are plotted in the positive side artificially.}
\end{figure}

In Fig.\ \ref{figure:flow2}, we show the incident energy dependence of
the calculated flows of nucleons and fragments with mass numbers
$A=1,2,3$ and 4 separately for $\argon+\almi$ collisions with the
fixed impact parameter 3 fm. For the moment we concentrate on the
results with Gogny force and the stochastic collision cross section
$1.0\sigma$, and the dependence on them will be discussed in a later
section. As the incident energy increases from 25 MeV/nucleon to 135
MeV/nucleon, the flow changes from negative (attractive) value to the
positive (repulsive) value at a certain incident energy called balance
energy.  The balance energy for nucleon flow is about 85 MeV/nucleon
and the balance energy for fragments is higher than it. In the energy
region where the flow is negative, the absolute value of the flow is
larger for heavier fragment. The flow takes its minimum (namely the
most attractive) value between 30 MeV/nucleon and 50 MeV/nucleon, and
at 25 MeV/nucleon the absolute value of flow is small. This is
evidently because of the almost spherical momentum distribution in
fusion-like events. These qualitative features as well as the
quantitative results are consistent to the experimental data in the
energy region 25 MeV $\le E/A\le$ 85 MeV where the data are available,
as is shown in Fig.\
\ref{figure:flowcomp}.  The calculated flow in this figure is
defined by Eq.\ (\ref{eq:flowdef}) in the similar way to the
experimental value. The experimental data are plotted in the positive
side because there is no way to determine the sign of the flow only
from the experiment, but evidently they should be considered to be
negative at least in the energy region $E/A<70$ MeV.

\begin{figure}
\ifx\epsfbox\undefined\else
\epsfxsize=\figsize
\centerline{\epsfbox{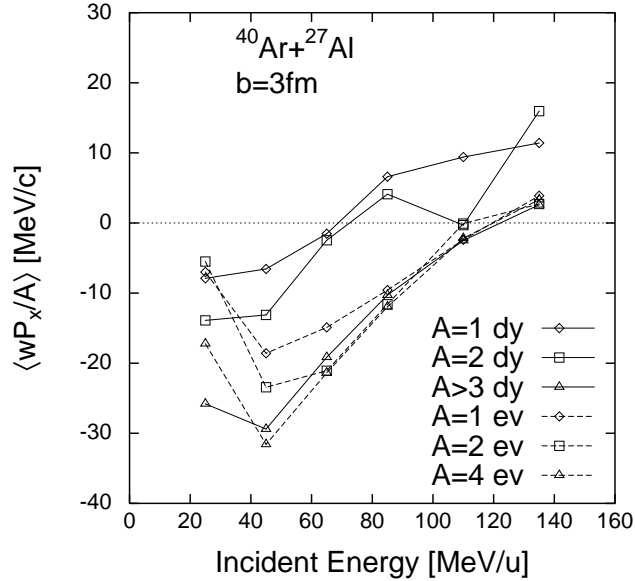}}
\fi
\caption{\label{figure:flow2-ana}
The flows of dynamically produced particles which are calculated
before the statistical decay process, and the flows of evaporated
particles which are calculated only from the decay products of the
statistical decay.
}
\end{figure}

In order to clarify the origin of the calculated feature of the flow,
especially the dependence of the flow on the particle mass number, we
classify the particles according to the time they are produced in the
calculation just as we have done in Ref.\ \CITE{ONOd}. Figure
\ref{figure:flow2-ana} shows the flow of dynamically produced particles
which existed at the end of the AMD calculation, namely at $t=t_{\rm
sw}$, and the flow of evaporated particles which are produced in the
statistical decay process. As we have found in the previous work for
$\carbon+\carbon$ collisions \cite{ONOd}, we can get the
interpretation that there are two components of flow at the end of the
dynamical stage of the reaction. The first component is the flow of
dynamically emitted nucleons and the second component is the flow of
excited fragments or the nuclear matter. In the energy region where
the flow is negative, the absolute value of the first component is
small because the nucleons are emitted by the stochastic two-nucleon
collisions which erase the effect of the attractive mean field.  The
second component has larger absolute value in which large effect of
the attractive interaction between projectile and target is remained.
Also in higher energy region, the flow of dynamically emitted nucleons
is more repulsive than the flow of excited fragments, which results in
the difference of the balance energies of nucleons and fragments in
the finally observed flow.

\section{Production Mechanism of Light Fragments and Their Flow}

\begin{figure}
\widetext
\ifx\epsfbox\undefined\else
\centerline{\epsfxsize=\figsize
\begin{minipage}{0.5\textwidth}
\epsfbox{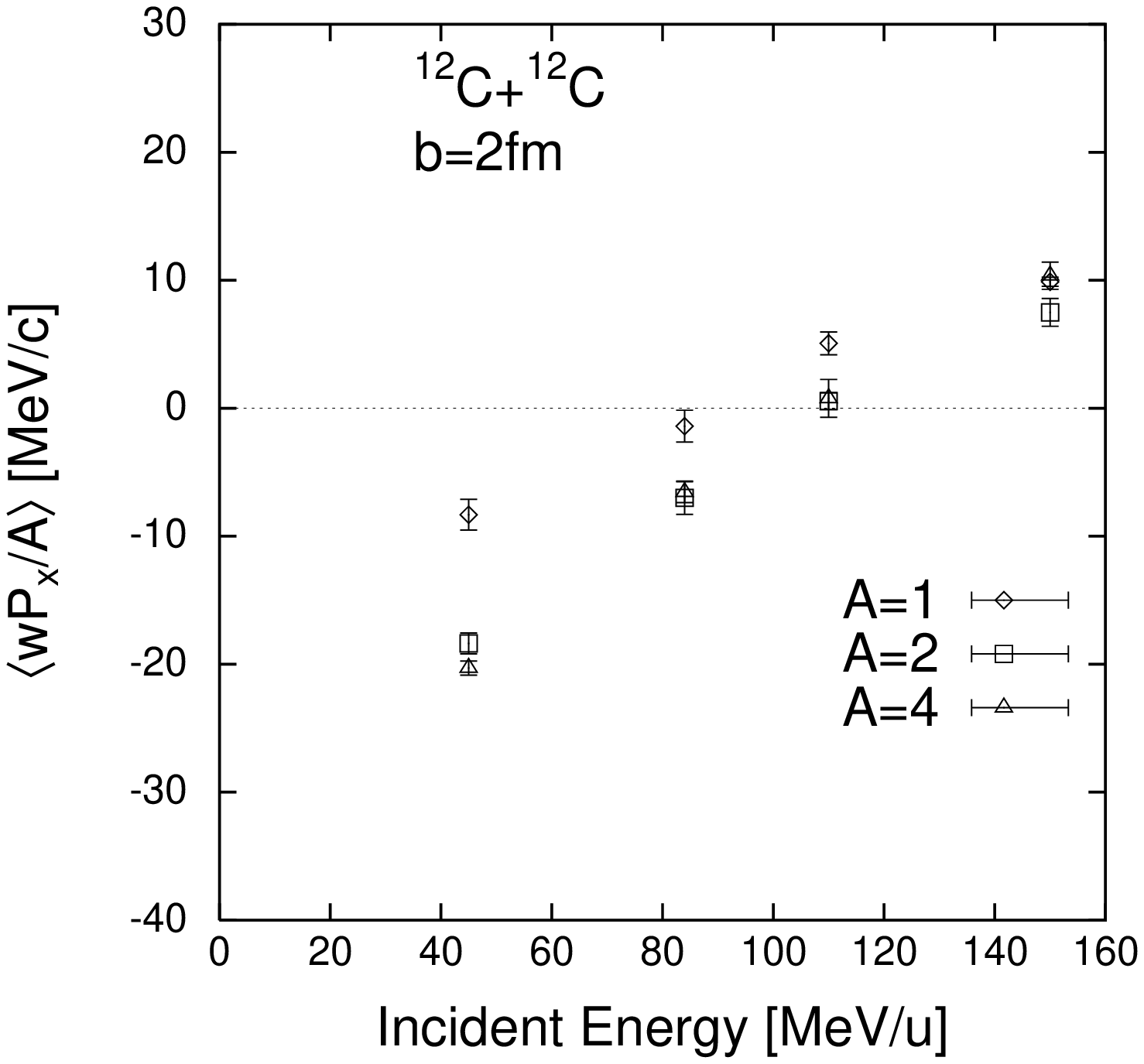}
\end{minipage}\hspace{-0.125\textwidth}
\begin{minipage}{0.5\textwidth}
\epsfbox{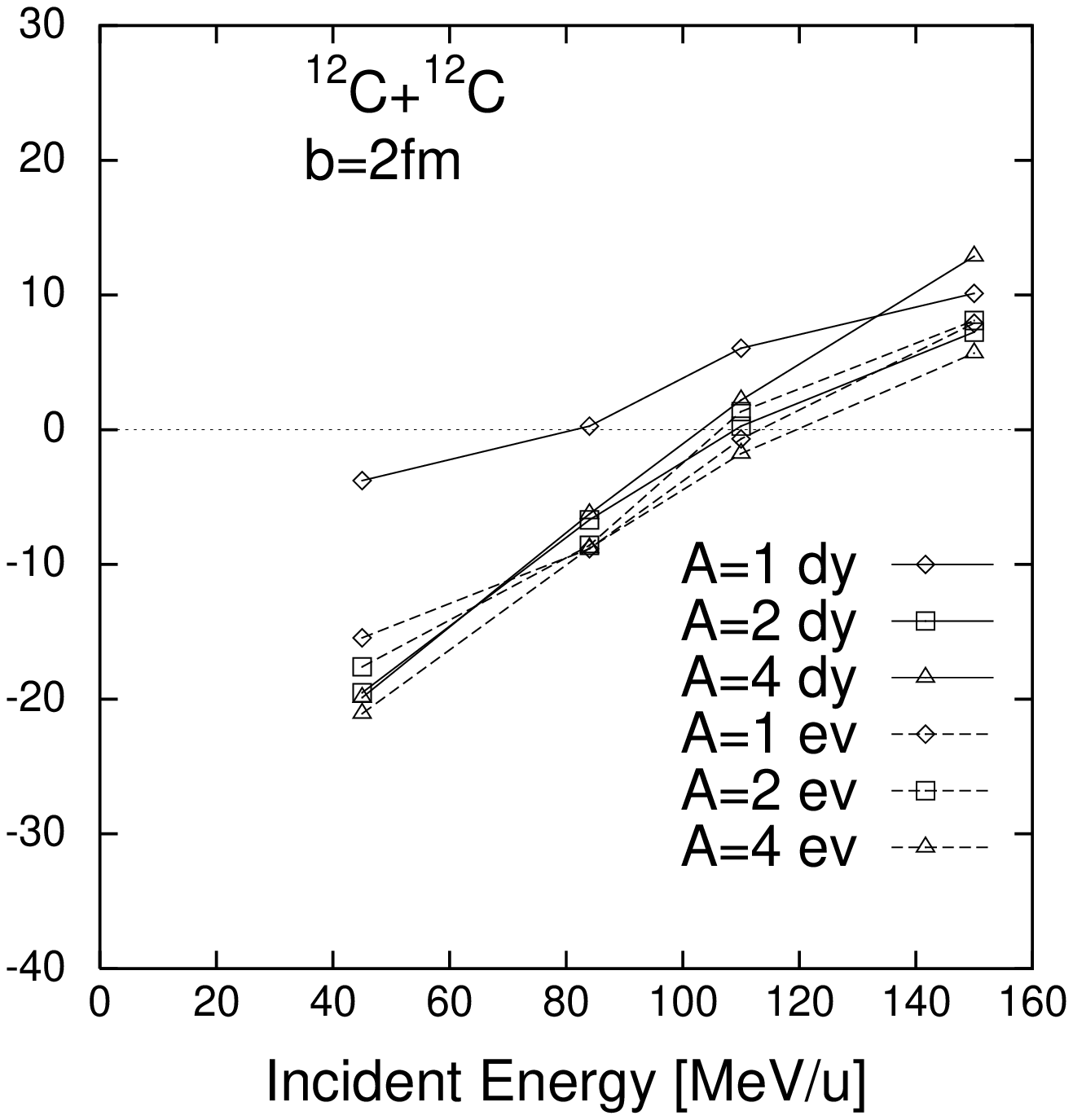}
\end{minipage}
}
\fi
\caption{\label{figure:CC-flow2}
Flow of nucleons, deuterons and $\alpha$ particles in
$\carbon+\carbon$ collisions with $b=2$ fm which was calculated in the
work of Ref.\ \protect\CITE{ONOd}.  }
\narrowtext
\end{figure}

Although we have found many similarities between $\argon+\almi$
collisions with $b=3$ fm and $\carbon+\carbon$ collisions with $b=2$
fm which we have studied in Ref.\ \CITE{ONOd}, there is an essential
difference in the magnitude of the flow with respect to the fragment
mass number.  In $\carbon+\carbon$ collisions the fragments with mass
numbers $A\ge2$ have almost an identical flow value as shown in Fig.\
\ref{figure:CC-flow2}.  On the other hand in $\argon+\almi$
collisions, the flows of particles with mass numbers $A=1,2,3$, and 4
are ordered according to their mass numbers as can be seen in Fig.\
\ref{figure:flow2}. This feature is also observed in experiments of
$\argon+{}^{45}{\rm Sc}$ collisions \cite{WESTFALL}. We have found
here that this difference of flow comes from the difference of the
flow behavior of dynamically produced deuterons while the flow of
evaporated deuterons has similar value to the flow of excited
fragments as shown in Fig.\ \ref{figure:flow2-ana}. The flow of
dynamically produced deuterons has a value which is close to the flow
of dynamical nucleons rather than the flow of heavier fragments in
$\argon+\almi$ collisions, while there is no difference between
deuteron flow and flow of heavier fragments in $\carbon+\carbon$
collisions as seen in Fig.\ \ref{figure:CC-flow2}.

\begin{table}
\caption{\label{table:dflowmech}
Comparison of the production mechanism of deuterons and $\alpha$
particles in $\carbon+\carbon$ collisions (at 45 MeV/nucleon and 84
MeV/nucleon) and $\argon+\almi$ collisions (at 45 MeV/nucleon and 85
MeV/nucleon). See text for detail.}
\begin{tabular}{cccccc}
& &\multispan2\hfill45 MeV/nucleon\hfill
&\multispan2\hfill84--85 MeV/nucleon\hfill\\
& &   C+C   &   Ar+Al  & C+C  & Ar+Al \\
\hline
$d$& $M$    &$  0.50       $&$  0.57$
            &$  1.04       $&$  0.86$\\
&  $P_{PT}$ &$ 17{\scriptstyle\pm2}\% $&$ 35{\scriptstyle\pm5}\% $
            &$ 15{\scriptstyle\pm1}\% $&$ 28{\scriptstyle\pm3}\% $\\
\hline
$\alpha$& $M$  &$  1.29       $&$  0.12  $
               &$  1.12       $&$  0.12  $\\
& $P_{PT}$     &$ 9{\scriptstyle\pm1}\% $&$ 35{\scriptstyle\pm10}\%$
               &$ 7{\scriptstyle\pm1}\% $&$ 21{\scriptstyle\pm8} \%$\\
\end{tabular}

\end{table}

The origin of this different flow behavior of the dynamically produced
deuterons in $\carbon+\carbon$ collisions and $\argon+\almi$
collisions can be understood in the following way by paying attention
to the production mechanism of deuterons.  We have traced back the
proton and the neutron of each dynamically produced deuteron in the
calculation and checked whether they originate from different initial
nuclei (i.e., one from the projectile and the other from the target)
or they come from the same nucleus (i.e., both from the projectile or
both from the target). The former probability is denoted by $P_{PT}$
and the latter probability is therefore $1-P_{PT}$.  The calculated
values of $P_{PT}$ are shown in Table \ref{table:dflowmech} together
with the multiplicities $M$ of deuterons and $\alpha$ particles. For
$\alpha$ particles, $P_{PT}$ is defined as the probability that the
$\alpha$ particle contains at least one nucleon from the projectile
and at least one nucleon from the target. In $\carbon+\carbon$
collisions only 15\% of dynamically produced deuterons are composed of
two nucleons from different nuclei, while this ratio is about 30\% in
$\argon+\almi$ collisions. With respect to $P_{PT}$, no significant
difference greater than the statistical error has been obtained
between 45 MeV/nucleon and 85 MeV/nucleon. Such deuterons should have
been produced by the coalescence, i.e., two nucleons which are emitted
by stochastic collisions have merged and formed deuterons when they
happened to be close in the phase space. It should be noted that the
$P_{PT}$ is at most about 50\% even if all deuterons are created by
the coalescence because two nucleons from the same nucleus can form a
deuteron by the coalescence with the similar probability to the two
nucleons from different nuclei. In other words, $P_{PT}=15\%$ in
$\carbon+\carbon$ collisions means that about 30\% of the dynamically
produced deuterons have been produced by the coalescence, and other
70\% of dynamical deuterons are produced by other mechanisms with
which only the nucleons from the same nucleus can form a deuteron.  In
the latter mechanisms, of course, the two nucleons have not suffered
the direct effect of the stochastic collisions. In the case of
$\argon+\almi$ collisions, $P_{PT}=30\%$ means that the coalescence
mechanism occupies about 60\% of the total deuterons produced
dynamically.  Strictly speaking, we should take account of the mass
asymmetry of the projectile and the target in the above discussion,
but this effect can be easily shown to be negligible in the case of
$\argon+\almi$ collisions.

Thus there is large dependence of the production mechanism of
dynamical deuterons on the mass number of the system.  This fact can
explain the difference of the deuteron flow between these reactions.
The deuterons created by the coalescence should reflect the momenta of
the nucleons which have been emitted by the stochastic collisions and
have happened to compose the deuterons, and therefore the flow of such
deuterons should have the flow similar to that of dynamically emitted
nucleons. This is the origin of the dynamical deuteron flow which is
close to the dynamical nucleon flow in $\argon+\almi$ collisions as
shown in Fig.\ \ref{figure:flow2-ana}. On the other hand, the
dynamical deuteron flow is the same as the flow of excited fragments
in $\carbon+\carbon$ collisions because most deuterons are created
without the direct effect of the stochastic collisions.

In Table \ref{table:dflowmech}, we can also find the significant
difference in the production mechanism of the dynamical $\alpha$
particles. There is a large difference in the calculated multiplicity,
the origin of which is the difference in the yield of $\alpha$
particles composed of four nucleons from the same nucleus. In
$\carbon+\carbon$ collisions, $P_{PT}$ is quite small and the
multiplicity of $\alpha$ particles which are produced without direct
effect of stochastic collisions is about 1, which is larger by an
order of magnitude than in $\argon+\almi$ collisions.

\section{Dependence of the Flow on $\bbox{\sigma}$ and EOS}
Although the ultimate purpose of the study of the flow is to determine
the EOS of the nuclear matter, we should also study the dependence of
the flow on the stochastic collision cross section ($\sigma$) at the
same time since there is much theoretical ambiguity in the adopted
stochastic collision cross section. Of course it is more desirable to
fix $\sigma$ by studying other quantities, such as the momentum
distribution of fragments discussed in Sec.\ III, and comparing
them to the experimental data. However, we will see in this section
fortunately that the study of the flow of nucleons and fragments gives
us much information on the EOS in spite of the uncertainty of
$\sigma$.

\begin{figure} 
\ifx\epsfbox\undefined\else
\centerline{\epsfxsize=\figsize\epsfbox{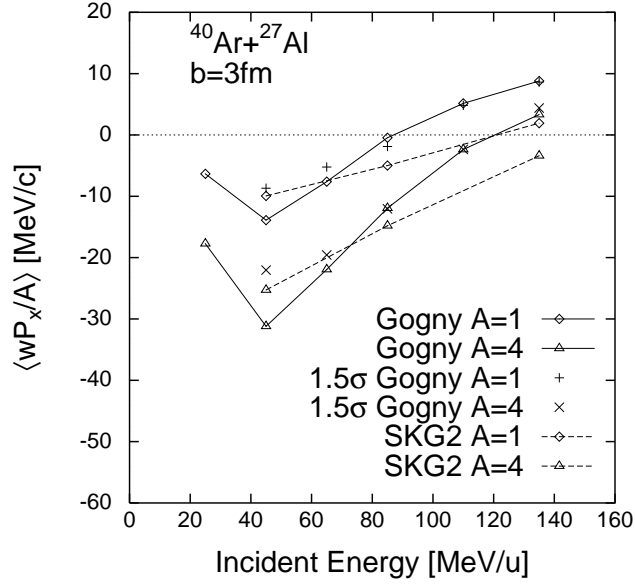}}
\fi
\caption{\label{figure:pxflow}
Nucleon flow and $\alpha$ flow in $\argon+\almi$ collisions with $b=3$
fm. Calculated results with Gogny force (solid line) and with SKG2
force (dashed line) are shown as well as the results with Gogny force
and increased cross section $1.5\sigma$ (pluses and crosses without
line).  }
\end{figure} 

\begin{figure} 
\begin{minipage}[t]{0.45\textwidth}
\ifx\epsfbox\undefined\else
\centerline{\hss\epsfxsize=\figsize\epsfbox{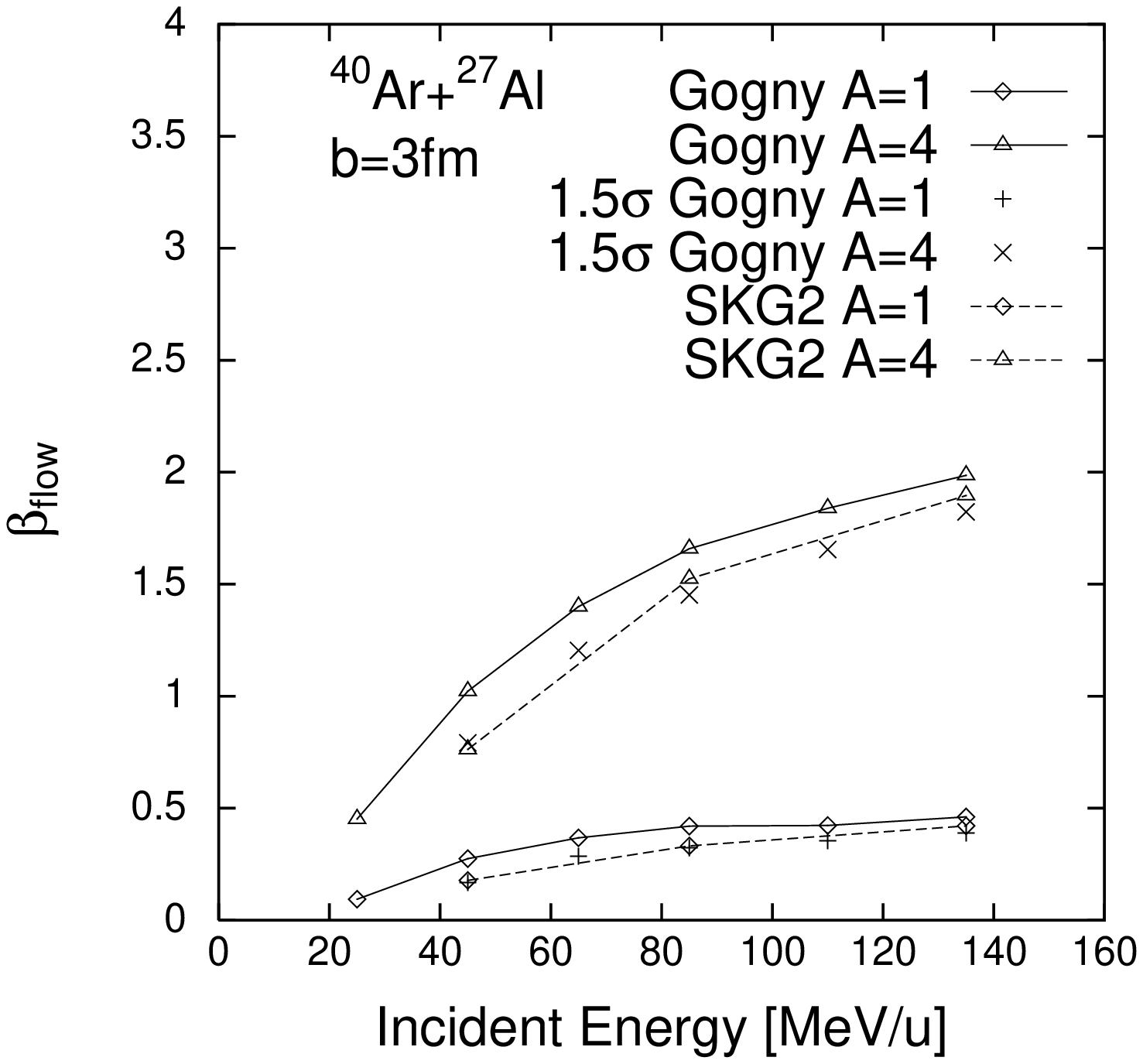}}
\fi
\caption{\label{figure:betaflow}
The quadrapole deformation parameter of the flow tensor.  }
\end{minipage}
\hspace{0.08\textwidth}
\begin{minipage}[t]{0.45\textwidth}
\ifx\epsfbox\undefined\else
\centerline{\hss\epsfxsize=\figsize\epsfbox{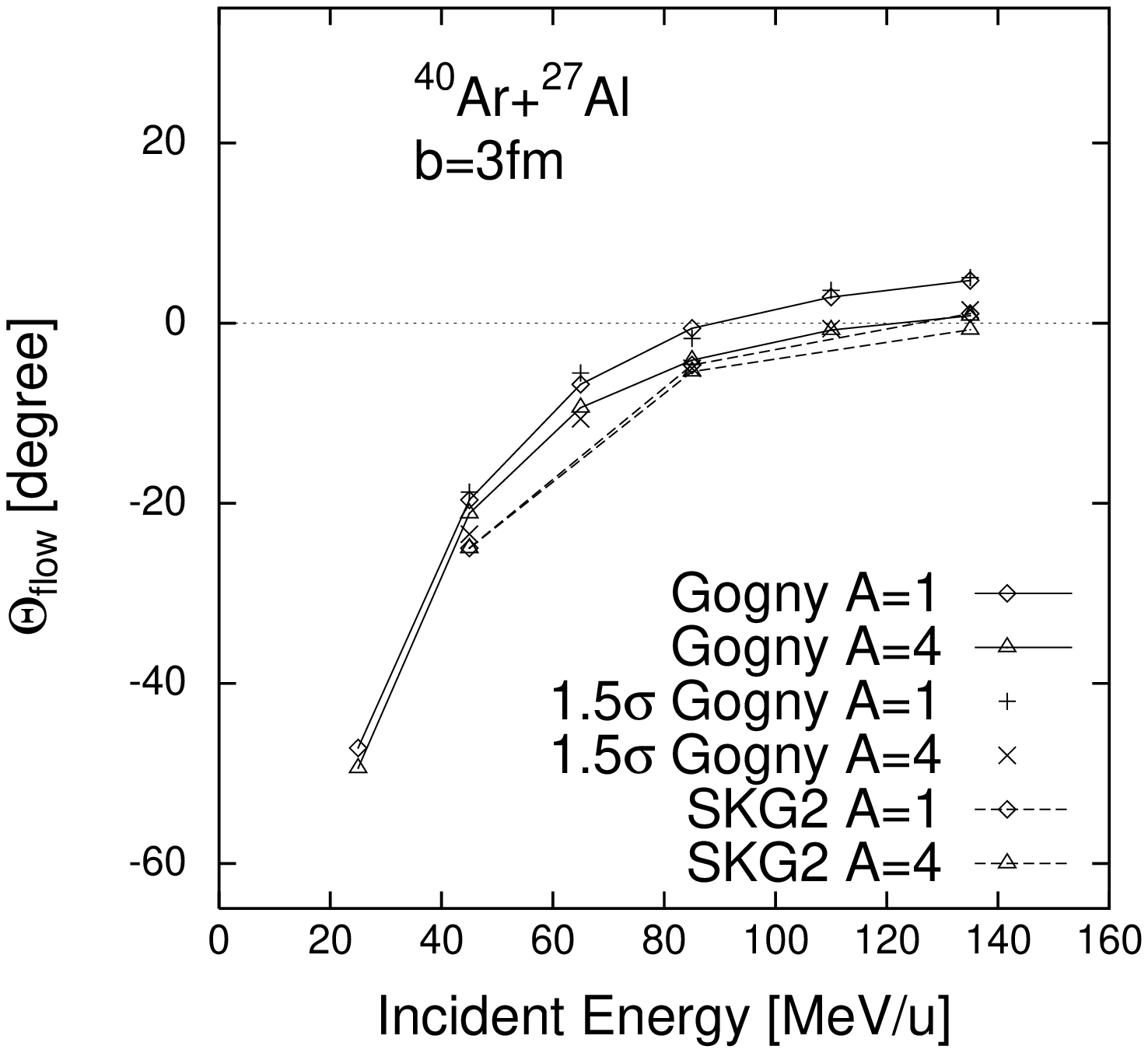}}
\fi
\caption{\label{figure:thetaflow}
The calculated value of the flow angle.  }
\end{minipage}
\end{figure} 

In Fig.\ \ref{figure:pxflow}, the flow of nucleons and $\alpha$
particles are shown for the calculation with the standard cross
section $1.0\sigma$ (symbols connected with solid lines) and increased
cross section $1.5\sigma$ (pluses and crosses without lines). Figs.\
\ref{figure:betaflow} and \ref{figure:thetaflow} show the deformation
parameter $\beta_{\rm flow}$ and the flow angle $\Theta_{\rm flow}$ of
the kinetic flow tensor
\begin{equation}
F_{ij}={\sum_{k}P_{ki} P_{kj}/A_k\over\sum_{k}A_k},
\end{equation}
where $k$ is the index of the produced fragments in all events and
$i,j=x,y,z$. The summation is limited to the fragments with a given
mass number in Figs.\ \ref{figure:betaflow} and
\ref{figure:thetaflow}. The flow angle $\Theta_{\rm flow}$ is the
angle between the beam direction and the eigenvector of $F_{ij}$
corresponding to the largest eigenvalue. Negative $\Theta_{\rm flow}$
means the attractive flow. Denoting the ratio of eigenvalues of
$F_{ij}$ as $e^{2t_1}:e^{2t_2}:e^{2t_3}$ with $t_1+t_2+t_3=0$ and
$t_1\le t_2\le t_3$, we define the deformation parameter $\beta_{\rm
flow}$ in the similar way to Eq.\ (\ref{eq:defbeta}).  The parameter
$\beta_{\rm flow}$ represents the degree of the dissipation of the
incident energy which can be also shown by the parallel momentum
distribution like in Fig.\ \ref{figure:pzdst}. Smaller $\beta_{\rm
flow}$ means larger dissipation. In relatively low energy region
$E/A\sim 45$ MeV/nucleon, the flow has large dependence on the
stochastic collision cross section as can be seen from Fig.\
\ref{figure:pxflow}. For larger $\sigma$ the absolute value of the
flow is smaller. As we have found in our previous work \cite{ONOd},
this decrease of the flow comes from the increase of the dissipated
components in the momentum distribution, which has appeared as the
decrease of $\beta_{\rm flow}$ in Fig.\ \ref{figure:betaflow}.  Note
that $\Theta_{\rm flow}$ is not sensitive to $\sigma$ because the
spherical dissipated component does not affect the eigenvectors of
$F_{ij}$.  The flow angle of $\alpha$ particles increases slightly
according to the increase of $\sigma$, but the decrease of $\beta_{\rm
flow}$ is much larger and, as the result, the value of the flow
decreases.  The flow angle $\Theta_{\rm flow}$ is also insensitive to
the mass number of the fragment, which is also an expected result from
the interpretation of the two component of flow. Namely, the
eigenvectors of nucleon flow tensor is essentially decided by the flow
of the evaporated nucleons which is identical to the flow of the
excited fragments, because the almost spherical component of
dynamically emitted nucleons does not affect them.  Therefore if the
independence of the flow angle is observed in experiment, it will be a
strong evidence of the two component of the flow in this energy
region.

\begin{figure} 
\ifx\epsfbox\undefined\else
\centerline{\epsfxsize=\figsize\epsfbox{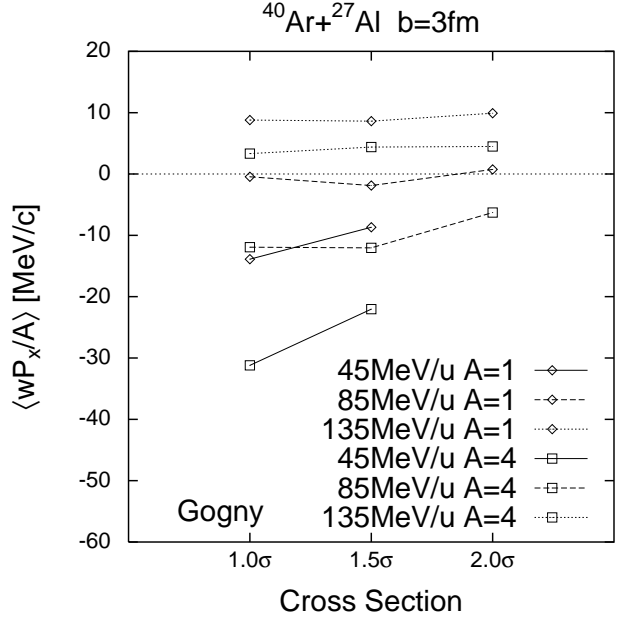}}
\fi
\caption{\label{figure:sigdep}
The $\sigma$-dependence of the nucleon flow and the $\alpha$ flow in
$\argon+\almi$ collisions with $b=3$ fm in the calculation with the
Gogny force.  }
\end{figure} 

In higher energy region $E/A\agt 85$ MeV, on the other hand, the
$\sigma$-dependence of the flow is quite small in our calculation.
The multiplicity of nucleons changes sensitively to $\sigma$ and the
dissipation increases to some degree for larger $\sigma$ as shown in
Fig.\ \ref{figure:betaflow}. However the change of the $\beta_{\rm
flow}$ is small compared to the absolute value of the deformation
parameter $\beta_{\rm flow}$.  The flow angle $\Theta_{\rm flow}$, and
therefore the value of flow, do not change at all.  When the cross
section is further increased up to $2.0\sigma$, a little
$\sigma$-dependence is found at 85 MeV/nucleon especially in the flow
of $\alpha$ particles, as shown in Fig.\ \ref{figure:sigdep}, because
the dissipation becomes rather large.  At 135 MeV/nucleon, however,
the $\sigma$-dependence is quite small in the wide region between
$1.0\sigma$ and $2.0\sigma$.

Not only with Gogny force which corresponds to the incompressibility
of the nuclear matter $K=228$ MeV, we have also made calculations with
the SKG2 force.  The SKG2 force corresponds to the large
incompressibility $K=373$ MeV and the mean field has no momentum
dependence.  The results with SKG2 force have been shown in Figs.
\ref{figure:pxflow}, \ref{figure:betaflow}, \ref{figure:thetaflow} and
\ref{figure:flowcomp} by dashed lines.

As is evident from these figures, the calculated flow with SKG2 force
does not reproduce the data of Refs.\ \cite{PETER,PETERb} in two
points. Firstly, the balance energy is too large even with this
effective interaction which corresponds to the stiff EOS.  Secondly,
the incident energy dependence of the flow of nucleons and fragments
is much smaller than with the Gogny force. The calculated results
suggest that the momentum dependence of the mean field is more
important than the density dependence in order to reproduce the large
incident energy dependence and the small balance energy.  Since the
$\sigma$-dependence of these features is small as we have discussed
above, we can conclude that the SKG2 force, i.e., the stiff EOS
without momentum dependence of the mean field is inconsistent to the
observed data of the flow of nucleons and fragments in the
intermediate energy region, while the Gogny force, which corresponds
to the soft EOS with momentum dependence of the mean field, reproduce
the data very well.

\begin{figure}
\widetext
\ifx\epsfbox\undefined\else
\centerline{\epsfxsize=\figsize
\begin{minipage}{0.5\textwidth}
\epsfbox{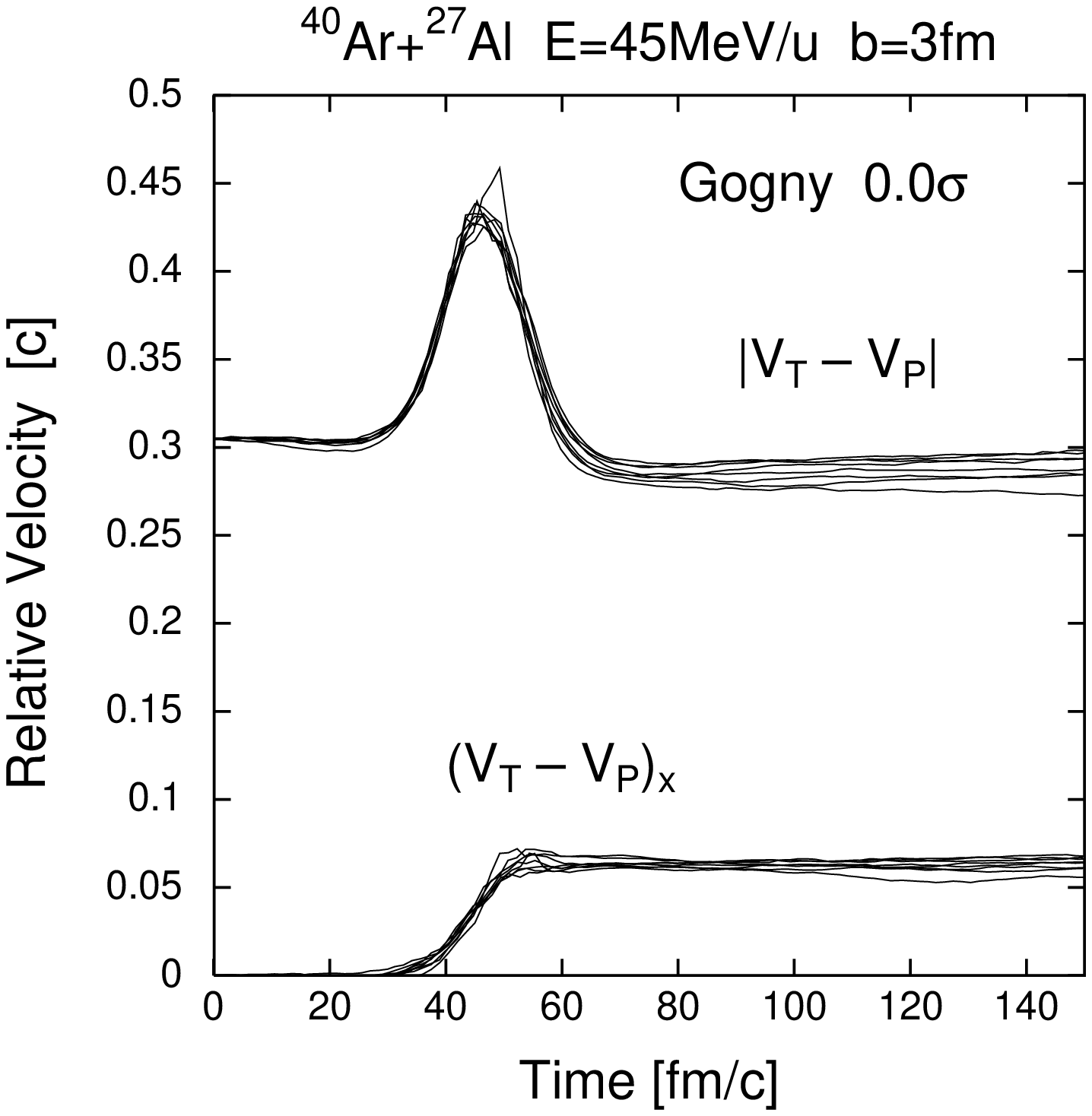}
\end{minipage}\hspace{-0.125\textwidth}
\begin{minipage}{0.5\textwidth}
\epsfbox{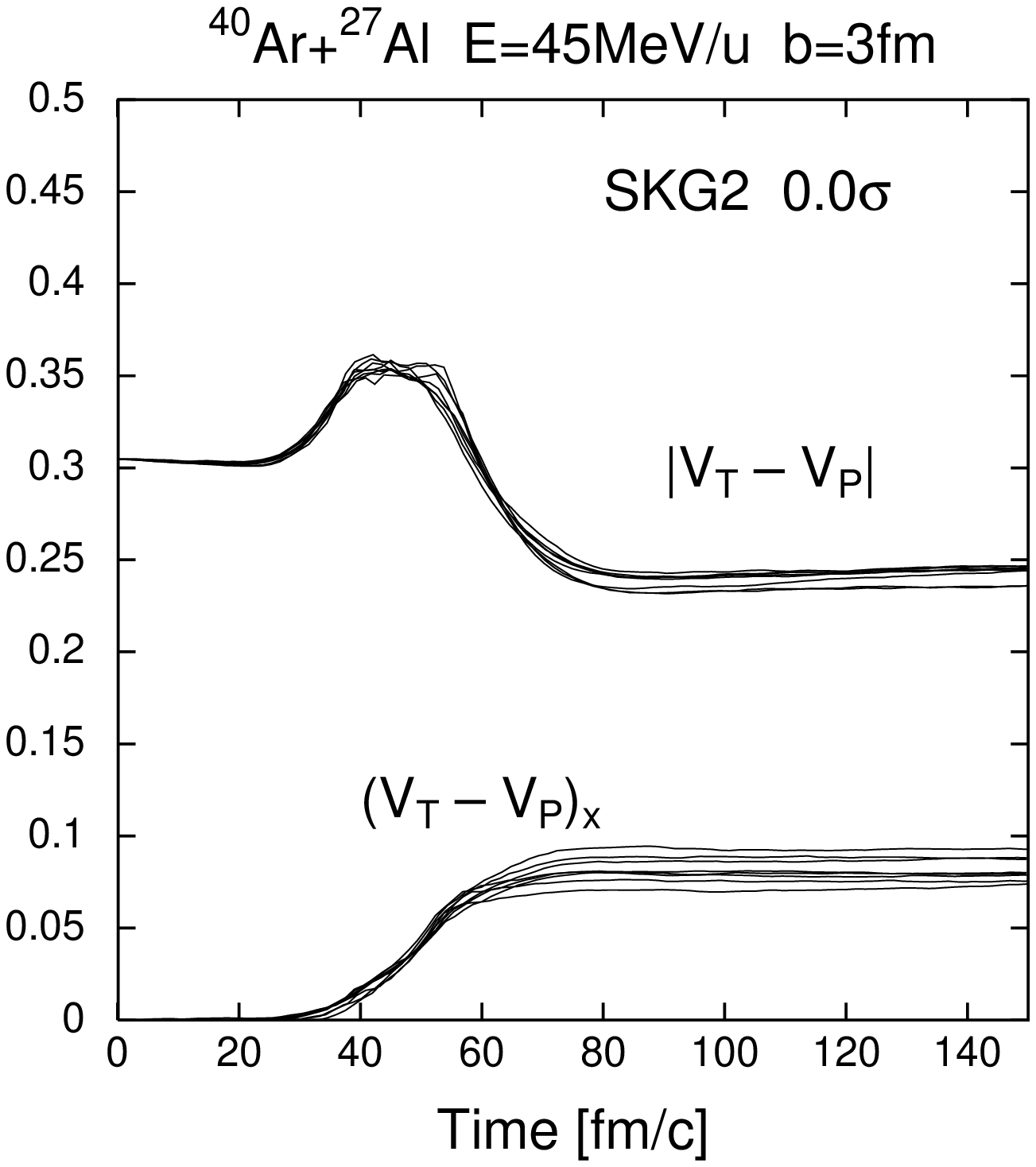}
\end{minipage}
}
\fi
\caption{\label{figure:nntraj}
The time development of the relative velocity of the target and the
projectile ${\vec V}_T-{\vec V}_P$ in the calculation without
stochastic collisions. Its absolute value and its $x$ component in 8
events at 45 MeV/nucleon are shown for the Gogny force (left) and the
SKG2 force (right).  }
\narrowtext
\end{figure}

Finally, we notice that the dissipation or the stopping is larger with
the SKG2 force than with the Gogny force when the same stochastic
collision cross section $1.0\sigma$ is adopted, as can be seen in
Fig.\ \ref{figure:betaflow}. In order to see whether this feature is
really due to the effect of the mean field, we calculated the
trajectory of the projectile and the target without any stochastic
collisions at the incident energy 45 MeV/nucleon and with the impact
parameter 3 fm. Figure \ref{figure:nntraj} shows the time development
of the absolute value and the $x$ component of the relative velocity
of the target and the projectile for 8 events. The difference among
events is due to the random orientation of the initial nuclei. The
velocity of the projectile (the target) is calculated from the time
development of the center-of-mass of the physical positions
$\{\mathop{\rm Re}{\vec W}_i\}$ defined by Eq.\ (\ref{eq:physcoord})
which originate from the projectile (the target).  It is clearly seen
that the acceleration due to the attractive mean field is suppressed
with the SKG2 force compared to the calculation with the Gogny force,
probably due to the large incompressibility of the SKG2 force
\cite{WADA}. As the result, with the SKG2 force, the interaction time
is longer and the one-body dissipation of the incident velocity is
larger. The final transverse relative velocity is larger with the SKG2
force, which is in accordance with the larger flow angle in Fig.\
\ref{figure:thetaflow} in the calculation with stochastic collisions.

\section{Summary}

In this paper, we studied $\argon+\almi$ collisions with AMD in the
incident energy region 25 MeV $\le E/A \le$ 135 MeV, aiming to study
the exclusive flow of nucleons and fragments and to determine the EOS
of the nuclear matter. The impact parameter was fixed to 3 fm in most
cases. The calculations were performed with two effective
interactions, namely the Gogny force which corresponds to the soft
EOS with reliable momentum dependence of the mean field and the SKG2
force which corresponds to the stiff EOS without momentum dependence
of the mean field. The calculation with the Gogny force reproduced
quite successfully many features of the experimental data of the flow
of nucleons and fragments, such as the balance energy, the absolute
value and the incident energy dependence of the flow, and the large
fragment flow compared to the nucleon flow. Such features were
explained by the concept of the two components of the flow, i.e., the
dynamical nucleon flow and the flow of excited fragments at the end of
the dynamical stage of the reaction.

On the other hand, the calculation with the SKG2 force failed in
reproducing the data. The calculated balance energy is too large and
the incident energy dependence of the flow is too small. We also
checked the dependence of the flow on the stochastic collision cross
section and found that there is some $\sigma$-dependence in the energy
region $E/A\sim45$ MeV which can be related to the dissipated
component in the momentum distribution but the $\sigma$-dependence is
quite small in the energy region $E/A\agt 85$ MeV. Therefore, even if
the theoretical ambiguity of the stochastic collision cross section in
the nuclear medium is taken into account, we can conclude that the
stiff EOS without the momentum dependence is inconsistent to the
observed data, while the data are well reproduced with the Gogny force
which gives a soft EOS with the momentum dependent mean field.

Our calculated results on the EOS depencence of the balance energy is
similar to the results of QMD by Ohnishi et al. \cite{OHNISHIconf}
where the statistical decay process was not taken into account.
However, our calculated $\sigma$-dependence is different form their
results.

We also compared the calculated results with the results for
$\carbon+\carbon$ collisions studied in our previous work \cite{ONOd}.
In addition to many similarities concerned with the flow of nucleons
and fragments, we found a significant difference in the flow of
dynamically produced deuterons.  In $\carbon+\carbon$ collisions, most
dynamical deuterons are produced without direct effect of the
stochastic collisions and their flow is identical to the flow of
excited fragments. On the other hand in $\argon+\almi$, many dynamical
deuterons are created by the coalescence of the nucleons which are
emitted by the stochastic collisions, and therefore the flow of
dynamical deuterons is close to the flow of dynamical nucleons.  In
the production mechanism of $\alpha$ particles, similar difference
were found between $\carbon+\carbon$ collisions and $\argon+\almi$
collisions.

\acknowledgements
The computational calculation for this work was partly supported by
Research Center for Nuclear Physics, Osaka University, as an RCNP
Computational Nuclear Physics Project (Project No.\ 92-B-04). Other
part of the computational calculation was performed on the super
computer FACOM VPP-500 in RIKEN. One of the authors (A.\ O.) is
supported by JSPS fellowship.

\end{document}